%% file: main.tex
\pdfoutput=1

\documentclass[12pt,a4paper]{article}

\usepackage{ifthen} 
\usepackage{rotating}
\newboolean{pdflatex}
\setboolean{pdflatex}{true} 

\newboolean{articletitles}
\setboolean{articletitles}{true} 

\newboolean{uprightparticles}
\setboolean{uprightparticles}{true} 

\newboolean{inbibliography}
\setboolean{inbibliography}{false} 

\input{preamble}

\begin{document}

\renewcommand{\thefootnote}{\fnsymbol{footnote}}
\setcounter{footnote}{1}

\input{title-LHCb-PAPER}

\renewcommand{\thefootnote}{\arabic{footnote}}
\setcounter{footnote}{0}



\pagestyle{plain} 
\setcounter{page}{1}
\pagenumbering{arabic}


%

\input{introduction}
\input{detector}
\input{selection}
\input{signal}

\input{multipion}
\input{effic}
\input{results}
\input{acknowledgements}



\addcontentsline{toc}{section}{References}
\setboolean{inbibliography}{true}
\bibliographystyle{LHCb}
\bibliography{main,LHCb-PAPER,LHCb-CONF,LHCb-DP,LHCb-TDR}

\newpage
\input{LHCb_Authorship_flat_02-Aug-2016}

\end{document}

%% file: preamble.tex
\usepackage[top=1in, bottom=1.25in, left=1in, right=1in]{geometry}

\usepackage{microtype}
\usepackage{lineno}  
\usepackage{xspace} 
\usepackage{caption} 

\usepackage{graphicx}  
\usepackage{color}
\usepackage{colortbl}
\graphicspath{{./figs/}} 

\usepackage{amsmath} 
\usepackage{amssymb}
\usepackage{amsfonts}
\usepackage{upgreek} 

\newcommand*\patchAmsMathEnvironmentForLineno[1]{%
\expandafter\let\csname old#1\expandafter\endcsname\csname #1\endcsname
\expandafter\let\csname oldend#1\expandafter\endcsname\csname
end#1\endcsname
 \renewenvironment{#1}%
   {\linenomath\csname old#1\endcsname}%
   {\csname oldend#1\endcsname\endlinenomath}%
}
\newcommand*\patchBothAmsMathEnvironmentsForLineno[1]{%
  \patchAmsMathEnvironmentForLineno{#1}%
  \patchAmsMathEnvironmentForLineno{#1*}%
}
\AtBeginDocument{%
\patchBothAmsMathEnvironmentsForLineno{equation}%
\patchBothAmsMathEnvironmentsForLineno{align}%
\patchBothAmsMathEnvironmentsForLineno{flalign}%
\patchBothAmsMathEnvironmentsForLineno{alignat}%
\patchBothAmsMathEnvironmentsForLineno{gather}%
\patchBothAmsMathEnvironmentsForLineno{multline}%
\patchBothAmsMathEnvironmentsForLineno{eqnarray}%
}

\usepackage{hyperref}    
\usepackage[all]{hypcap} 

\input{lhcb-symbols-def} 

\usepackage{cite} 
\usepackage{mciteplus}

%% file: lhcb-symbols-def.tex
\usepackage{xspace} 
\usepackage{upgreek}

\def\lhcb {\mbox{LHCb}\xspace}

\def\MagUp {\mbox{\em Mag\kern -0.05em Up}\xspace}

\ifthenelse{\boolean{uprightparticles}}%
{

 \def\Pmu         {\ensuremath{\upmu}\xspace}

 \def\Ppi         {\ensuremath{\uppi}\xspace}                 
                  
 \def\Prho        {\ensuremath{\uprho}\xspace}

 \def\Ppsi        {\ensuremath{\uppsi}\xspace}

 \def\PDelta      {\ensuremath{\Delta}\xspace}                 
 \def\PXi      {\ensuremath{\Xi}\xspace}                 
 \def\PLambda      {\ensuremath{\Lambda}\xspace}                 
 \def\PSigma      {\ensuremath{\Sigma}\xspace}                 
 \def\POmega      {\ensuremath{\Omega}\xspace}                 
 \def\PUpsilon      {\ensuremath{\Upsilon}\xspace}                 
 
 \def\PB      {\ensuremath{\mathrm{B}}\xspace}                 
                  
 \def\PD      {\ensuremath{\mathrm{D}}\xspace}

 \def\PJ      {\ensuremath{\mathrm{J}}\xspace}                 
 \def\PK      {\ensuremath{\mathrm{K}}\xspace}

 \def\PW      {\ensuremath{\mathrm{W}}\xspace}

 \def\Pb      {\ensuremath{\mathrm{b}}\xspace}                 
 \def\Pc      {\ensuremath{\mathrm{c}}\xspace}

 \def\Pi      {\ensuremath{\mathrm{i}}\xspace}

 \def\Pp      {\ensuremath{\mathrm{p}}\xspace}

 \def\Pu      {\ensuremath{\mathrm{u}}\xspace}

}
{

 \def\Pmu         {\ensuremath{\mu}\xspace}

 \def\Ppi         {\ensuremath{\pi}\xspace}                 
                  
 \def\Prho        {\ensuremath{\rho}\xspace}

 \def\Ppsi        {\ensuremath{\psi}\xspace}                 
                  
 \mathchardef\PDelta="7101
 \mathchardef\PXi="7104
 \mathchardef\PLambda="7103
 \mathchardef\PSigma="7106
 \mathchardef\POmega="710A
 \mathchardef\PUpsilon="7107
                  
 \def\PB      {\ensuremath{B}\xspace}                 
                  
 \def\PD      {\ensuremath{D}\xspace}

 \def\PJ      {\ensuremath{J}\xspace}                 
 \def\PK      {\ensuremath{K}\xspace}

 \def\PW      {\ensuremath{W}\xspace}

 \def\Pb      {\ensuremath{b}\xspace}                 
 \def\Pc      {\ensuremath{c}\xspace}

 \def\Pi      {\ensuremath{i}\xspace}

 \def\Pp      {\ensuremath{p}\xspace}

 \def\Pu      {\ensuremath{u}\xspace}

}

\makeatletter
\ifcase \@ptsize \relax
  \newcommand{\miniscule}{\@setfontsize\miniscule{4}{5}}
\or
  \newcommand{\miniscule}{\@setfontsize\miniscule{5}{6}}
\or
  \newcommand{\miniscule}{\@setfontsize\miniscule{5}{6}}
\fi
\makeatother

\DeclareRobustCommand{\optbar}[1]{\shortstack{{\miniscule (\rule[.5ex]{1.25em}{.18mm})}
  \\ [-.7ex] $#1$}}

\def\mup        {{\ensuremath{\Pmu^+}}\xspace}
\def\mun        {{\ensuremath{\Pmu^-}}\xspace} 
\def\mumu       {{\ensuremath{\Pmu^+\Pmu^-}}\xspace}

\def\uquark    {{\ensuremath{\Pu}}\xspace}

\def\cquark    {{\ensuremath{\Pc}}\xspace}

\def\bquark    {{\ensuremath{\Pb}}\xspace}
\def\bquarkbar {{\ensuremath{\overline \bquark}}\xspace}

\def\pion   {{\ensuremath{\Ppi}}\xspace}

\def\pip    {{\ensuremath{\pion^+}}\xspace}
\def\pim    {{\ensuremath{\pion^-}}\xspace}

\def\kaon    {{\ensuremath{\PK}}\xspace}
  \def\Kbar    {{\kern 0.2em\overline{\kern -0.2em \PK}{}}\xspace}

\def\KorKbar    {\kern 0.18em\optbar{\kern -0.18em K}{}\xspace}

\def\Kp      {{\ensuremath{\kaon^+}}\xspace}
\def\Km      {{\ensuremath{\kaon^-}}\xspace}

  \def\Dbar    {{\kern 0.2em\overline{\kern -0.2em \PD}{}}\xspace}
\def\D       {{\ensuremath{\PD}}\xspace}

\def\DorDbar    {\kern 0.18em\optbar{\kern -0.18em D}{}\xspace}
\def\Dz      {{\ensuremath{\D^0}}\xspace}

\def\Dstarp  {{\ensuremath{\D^{*+}}}\xspace}

\def\B       {{\ensuremath{\PB}}\xspace}
\def\Bbar    {{\ensuremath{\kern 0.18em\overline{\kern -0.18em \PB}{}}}\xspace}

\def\BorBbar    {\kern 0.18em\optbar{\kern -0.18em B}{}\xspace}

\def\Bu      {{\ensuremath{\B^+}}\xspace}

\def\Bp      {{\ensuremath{\Bu}}\xspace}

\def\Bc      {{\ensuremath{\B_\cquark^+}}\xspace}

\def\jpsi     {{\ensuremath{{\PJ\mskip -3mu/\mskip -2mu\Ppsi\mskip 2mu}}}\xspace}
\def\psitwos  {{\ensuremath{\Ppsi{(2\mathrm{S})}}}\xspace}

  \def\Y#1S{\ensuremath{\PUpsilon{(#1S)}}\xspace}

\def\proton      {{\ensuremath{\Pp}}\xspace}

\def\Lbar        {{\ensuremath{\kern 0.1em\overline{\kern -0.1em\PLambda}}}\xspace}
\def\LorLbar    {\kern 0.18em\optbar{\kern -0.18em \PLambda}{}\xspace}

\def\BF         {{\ensuremath{\mathcal{B}}}\xspace}

\def\BR         {\BF}

\def\to                 {\ensuremath{\rightarrow}\xspace}

\def\AT#1     {\ensuremath{A_{\mathrm{T}}^{#1}}\xspace}           

\def\C#1      {\ensuremath{\mathcal{C}_{#1}}\xspace}                       
\def\Cp#1     {\ensuremath{\mathcal{C}_{#1}^{'}}\xspace}                    
\def\Ceff#1   {\ensuremath{\mathcal{C}_{#1}^{\mathrm{(eff)}}}\xspace}        
\def\Cpeff#1  {\ensuremath{\mathcal{C}_{#1}^{'\mathrm{(eff)}}}\xspace}       
\def\Ope#1    {\ensuremath{\mathcal{O}_{#1}}\xspace}                       
\def\Opep#1   {\ensuremath{\mathcal{O}_{#1}^{'}}\xspace}                    

\newcommand{\tev}{\ifthenelse{\boolean{inbibliography}}{\ensuremath{~T\kern -0.05em eV}\xspace}{\ensuremath{\mathrm{\,Te\kern -0.1em V}}}\xspace}
\newcommand{\gev}{\ensuremath{\mathrm{\,Ge\kern -0.1em V}}\xspace}
\newcommand{\mev}{\ensuremath{\mathrm{\,Me\kern -0.1em V}}\xspace}
\newcommand{\kev}{\ensuremath{\mathrm{\,ke\kern -0.1em V}}\xspace}
\newcommand{\ev}{\ensuremath{\mathrm{\,e\kern -0.1em V}}\xspace}
\newcommand{\gevc}{\ensuremath{{\mathrm{\,Ge\kern -0.1em V\!/}c}}\xspace}
\newcommand{\mevc}{\ensuremath{{\mathrm{\,Me\kern -0.1em V\!/}c}}\xspace}
\newcommand{\gevcc}{\ensuremath{{\mathrm{\,Ge\kern -0.1em V\!/}c^2}}\xspace}
\newcommand{\gevgevcccc}{\ensuremath{{\mathrm{\,Ge\kern -0.1em V^2\!/}c^4}}\xspace}
\newcommand{\mevcc}{\ensuremath{{\mathrm{\,Me\kern -0.1em V\!/}c^2}}\xspace}

\def\mum  {\ensuremath{{\,\upmu\mathrm{m}}}\xspace}

\def\invfb   {\ensuremath{\mbox{\,fb}^{-1}}\xspace}

\newcommand{\chisq}{\ensuremath{\chi^2}\xspace}

\newcommand{\chisqip}{\ensuremath{\chi^2_{\text{IP}}}\xspace}

\def\gsim{{~\raise.15em\hbox{$>$}\kern-.85em
          \lower.35em\hbox{$\sim$}~}\xspace}
\def\lsim{{~\raise.15em\hbox{$<$}\kern-.85em
          \lower.35em\hbox{$\sim$}~}\xspace}

\def\ptot       {\mbox{$p$}\xspace}
\def\pt         {\mbox{$p_{\mathrm{ T}}$}\xspace}

\def\evtgen     {\mbox{\textsc{EvtGen}}\xspace}

\def\geant      {\mbox{\textsc{Geant4}}\xspace}

\def\photos     {\mbox{\textsc{Photos}}\xspace}

\def\pythia     {\mbox{\textsc{Pythia}}\xspace}

\def\tell1  {TELL1\xspace}
\def\ukl1   {UKL1\xspace}

\def\btofivepi    {\mbox{$\Bu\to\jpsi3\pip2\pim$}\xspace}
\def\fivepi       {\mbox{$\jpsi3\pip2\pim$}\xspace}
\def\btopsipi     {\mbox{$\Bu\to\psitwos[\to\jpsi\pip\pim]\pip\pip\pim$}\xspace}
\def\btopsipis    {\mbox{$\Bu\to\psitwos\pip\pip\pim$}\xspace}

\def\btopsik      {\mbox{$\Bu\to\psitwos[\to\jpsi\pip\pim]\Kp$}\xspace}
\def\btopsiks     {\mbox{$\Bu\to\psitwos\Kp$}\xspace}

\def\psitomu      {\mbox{$\jpsi\to\mumu$}\xspace}
\def\psitwostopsi {\mbox{$\psitwos\to\jpsi\pip\pim$}\xspace}
\def\jpsipipi     {\mbox{$\jpsi\pip\pim$}\xspace}


%% file: title-LHCb-PAPER.tex

\begin{titlepage}
\pagenumbering{roman}

\vspace*{-1.5cm}
\centerline{\large EUROPEAN ORGANIZATION FOR NUCLEAR RESEARCH (CERN)}
\vspace*{1.5cm}
\noindent
\begin{tabular*}{\linewidth}{lc@{\extracolsep{\fill}}r@{\extracolsep{0pt}}}
\ifthenelse{\boolean{pdflatex}}
{\vspace*{-2.7cm}\mbox{\!\!\!\includegraphics[width=.14\textwidth]{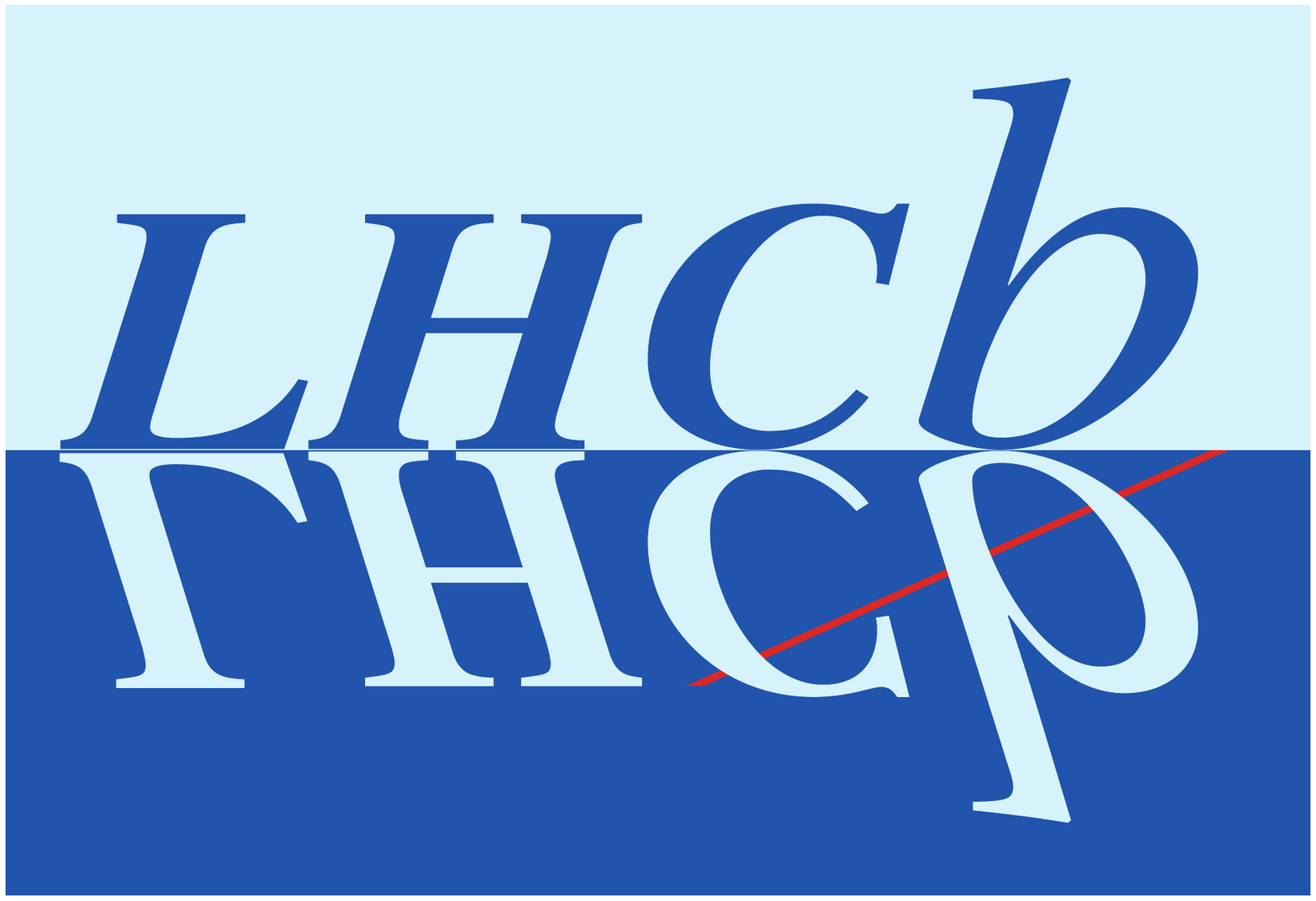}} & &}%
{\vspace*{-1.2cm}\mbox{\!\!\!\includegraphics[width=.12\textwidth]{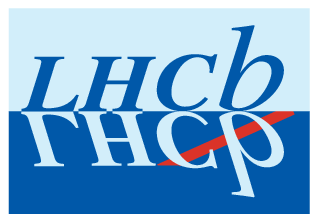}} & &}%
\\
 & & CERN-EP-2016-247 \\  
 & & LHCb-PAPER-2016-040 \\  
 & & October 5, 1016 \\ 
 & & 
\end{tabular*}

\vspace*{4.0cm}

{\normalfont\bfseries\boldmath\huge
  \begin{center}
    Observation of \btofivepi and \btopsipis decays
  \end{center}
}

\vspace*{2.0cm}

\begin{center}
The LHCb collaboration\footnote{Authors are listed at the end of this paper.}
\end{center}

\vspace{\fill}

\begin{abstract}
\noindent The decays \btofivepi and \btopsipis are observed for the~first time using 
a~data sample corresponding 
to an~integrated luminosity of~$3.0\invfb$, collected by 
the~\lhcb experiment in proton-proton collisions at 
the~centre\nobreakdash-of\nobreakdash-mass energies 
of~7 and 8\tev. 
The~branching fractions relative to that of \btopsiks are measured to be
\begin{eqnarray*}
  \dfrac {  \BR\left( \btofivepi \right)}
  {  \BR\left( \btopsiks \right) } 
  &=& 
  \left( 1.88\pm0.17\pm0.09\right)\times10^{-2}~, \\
  \dfrac {  \BR\left( \btopsipis \right)}
  {  \BR\left( \btopsiks \right) } 
  &=& 
  \left( 3.04\pm0.50\pm0.26\right)\times10^{-2}~,
\end{eqnarray*}
where the first uncertainties are statistical and the second are systematic.
\end{abstract}

\vspace*{2.0cm}

\begin{center}
  Published in Eur.~Phys.~J.~C~(2017)~77:72.
\end{center}

\vspace{\fill}

{\footnotesize 
  \centerline{\copyright~CERN on behalf of the \lhcb collaboration, 
    licence \href{http://creativecommons.org/licenses/by/4.0/}{CC-BY-4.0}.}}
\vspace*{2mm}

\end{titlepage}


\newpage
\setcounter{page}{2}
\mbox{~}

\cleardoublepage

%% file: introduction.tex
\section{Introduction}
\label{sec:Introduction}

The~\Bu~meson is a bound state of a heavy $\bquarkbar$~quark and a $\uquark$~quark, with 
well known properties and a large number of decay modes~\cite{PDG}, 
but little is known about decays of \Bp~mesons to a~\jpsi~meson plus
a~large number of light hadrons. 
The~\mbox{$\Bu\to\jpsi3\pip2\pim$}~decay channel is of particular interest, 
since it is one of the~highest multiplicity final states currently experimentally accessible. 
Evidence for the corresponding decay of the~\Bc~meson has recently been reported by 
the~LHCb collaboration~\cite{LHCb-PAPER-2014-009},
with the~measured branching fraction 
and qualitative behaviour of the multipion system consistent 
with expectations from QCD factorisation~\cite{Bauer:1986bm,Wirbel:1988ft}. 
In~this~scheme, the~\mbox{$\Bc\to\jpsi3\pip2\pim$}~decay 
is characterized  by the~form factors of the~\mbox{$\Bc\to\jpsi\PW^+$} transition and 
the~spectral functions for the~conversion of the~$\PW^+$~boson into 
light hadrons~\cite{Lesha,Likhoded:2013iua,Berezhnoy:2011is,Likhoded:2009ib}. 
Different decay topologies contribute to decays of \Bu~mesons 
into charmonia and light hadrons, affecting  
the~dynamics of the multipion system and enabling  
the~role of factorisation in \Bu~meson decays to be probed.

This~paper describes an~analysis of \btofivepi~decays, 
including decays to the~same final state that proceed through an~intermediate \psitwos~resonance. 
Charge\nobreakdash-conjugate modes are implied throughout the paper.  
The~ratios of the branching fractions for each of these decays to that of the~normalisation decay
$\Bu\to\psitwos\Kp$,
\begin{align}
  \begin{split}
    R_{5\pi}    &\equiv\frac{\BR(\btofivepi)}{\BR(\btopsiks)}~,     \\ 
    R_{\psitwos} &\equiv\frac{\BR(\btopsipis)}{\BR(\btopsiks)}~,
  \end{split}
  \label{eq:rate_fivepi}
\end{align}
are measured, 
where the $\psitwos$~meson is reconstructed in the $\jpsi\pip\pim$~final state and the 
$\jpsi$~meson~is reconstructed in its dimuon decay channel.
In~addition, a~search for intermediate resonances 
in the~multipion system is performed
and a~phase\nobreakdash-space model is compared to the~data and to 
the~predictions from QCD factorisation~\cite
{Bauer:1986bm,Wirbel:1988ft,Lesha,Likhoded:2013iua,Berezhnoy:2011is,Likhoded:2009ib}.
The~results are based on $\proton\proton$~collision data corresponding to an~integrated 
luminosity of $1.0\invfb$ and $2.0\invfb$ collected by the~\lhcb experiment 
at centre\nobreakdash-of\nobreakdash-mass energies of $\sqrt{s}=7\tev$ and $8\tev$, respectively.

%% file: detector.tex
\section{Detector and simulation}
\label{sec:Detector}

The~\lhcb detector~\cite{Alves:2008zz,LHCb-DP-2014-002} is a single-arm forward spectrometer 
covering the~pseudorapidity range~\mbox{$2<\eta<5$}, designed for the~study of particles 
containing \bquark or \cquark~quarks. 
The~detector includes a~high-precision tracking system consisting of 
a~silicon-strip vertex detector surrounding the~$\proton\proton$~interaction region, 
a~large-area silicon-strip detector located upstream of a~dipole magnet 
with a~bending power of about~$4{\mathrm{\,Tm}}$, 
and three stations of silicon\nobreakdash-strip detectors and straw drift tubes placed 
downstream of the~magnet. 
The~tracking system provides a~measurement of momentum, \ptot, 
of charged particles with a~relative uncertainty that varies from~0.5\% 
at low momentum to~1.0\% at~200\gevc. 
The~minimum distance of a~track to a~primary vertex\,(PV), 
the~impact parameter, is measured with 
a~resolution of~\mbox{$(15+29/\pt)\mum$}, 
where~\pt is the~component of the~momentum transverse to the~beam in\,\gevc. 
Different types of charged hadrons are distinguished using information from 
two ring\nobreakdash-imaging Cherenkov detectors\,(RICH). 
Photons, electrons and hadrons are identified by a~calorimeter system consisting of 
scintillating\nobreakdash-pad and preshower detectors, 
an~electromagnetic calorimeter and a~hadronic calorimeter. 
Muons are identified by a~system composed of alternating layers 
of iron and multiwire proportional chambers. 

The~online event selection is performed by a~trigger~\cite{LHCb-DP-2012-004}, 
which consists of a~hardware stage, based on information from 
the~calorimeter and muon systems, followed by a~software stage, 
which applies a~full event reconstruction.
The~hardware trigger selects 
muon candidates with $\pt>1.48\,(1.76)\gevc$ or
pairs of opposite-sign muon candidates 
with a~requirement that the~product of the~muon transverse momenta 
is larger than $1.7\,(2.6)\,\mathrm{GeV}^2/c^2$ 
for data collected at $\sqrt{s}=7\,(8)\tev$.
The subsequent software trigger is composed of two stages, 
the~first of which performs a~partial event reconstruction,
while full event reconstruction is done at the~second stage.
In the~software trigger 
the~invariant mass of well\nobreakdash-reconstructed pairs of 
oppositely charged muons forming a~good\nobreakdash-quality
two\nobreakdash-track vertex is required to exceed 2.7\gevcc,
and the~two\nobreakdash-track vertex 
is required to be significantly displaced from all PVs.

The~analysis technique reported below is validated using simulated events. 
In~the~simulation, $\proton\proton$~collisions are generated using
\pythia~\cite{Sjostrand:2006za,*Sjostrand:2007gs} with a~specific \lhcb configuration~\cite{LHCb-PROC-2010-056}. 
Decays of hadronic particles are described by \evtgen~\cite{Lange:2001uf}, 
in which final\nobreakdash-state radiation is generated using \photos~\cite{Golonka:2005pn}. 
A~model assuming QCD~factorisation
is implemented to generate the~decays $\btofivepi$ and $\btopsipis$~\cite{Lesha}.
The~interaction of the~generated particles with the~detector and its response 
are implemented using the~\geant toolkit~\cite{Allison:2006ve, *Agostinelli:2002hh} as described in Ref.~\cite{LHCb-PROC-2011-006}.

%% file: selection.tex
\section{Candidate selection}
\label{sec:selection}

The~decays \btofivepi, \btopsipis and \btopsiks are reconstructed using the decay modes \psitomu and \psitwostopsi followed by \psitomu.
Similar selection criteria are applied to all channels in order to minimize the systematic uncertainties.

Muon, pion and kaon candidates are selected from well\nobreakdash-reconstructed tracks
and are identified using information from the~RICH, calorimeter and muon detectors. 
Muon candidates are required to have a~transverse momentum larger than~$550\mevc$. 
Both pion and kaon candidates are required to have a~transverse momentum larger than~$250\mevc$ 
and momentum between~$3.2\gevc$ and~$150\gevc$ to allow good particle identification. 
To reduce combinatorial background due to tracks from the $\proton\proton$~interaction vertex, 
only tracks that are inconsistent with originating from a PV are used. 

Pairs of oppositely charged muons originating from a~common vertex are combined to 
form $\jpsi\to\mumu$~candidates. The~mass of the~dimuon combination is required to be 
between $3.020$ and $3.135\gevcc$. The~asymmetric mass range around 
the~known $\jpsi$~meson mass~\cite{PDG} 
is chosen to include the low\nobreakdash-mass tail due to final\nobreakdash-state radiation.

To~form a~\Bp~candidate, the~selected~\jpsi candidates are combined with
$3\pip2\pim$ or  $\Kp\pip\pim$~candidates for the~signal and control decays, respectively. 
Each~\Bp~candidate is associated with the~PV with respect to which it has the smallest~\chisqip, 
which is defined as the~difference in the~vertex fit~\chisq of the~PV with and without 
the~particle under consideration. 
To~improve the~mass resolution, a~kinematic fit~\cite{Hulsbergen:2005pu} is applied.
In~this fit the~mass of the~$\mup\mun$~combination is fixed to the~known \jpsi mass,  
and the~\Bp~candidate's momentum vector is required to originate at the~associated PV. 
A~good-quality fit is required to further suppress combinatorial background. 
In~addition, the~measured decay time of the~\Bp candidate, calculated with respect
to the~associated PV, is required to be larger than~$200\mum/c$, to suppress 
background from particles coming from the~PV.

%% file: signal.tex
\section{Signal and normalisation yields}
\label{sec:signals}

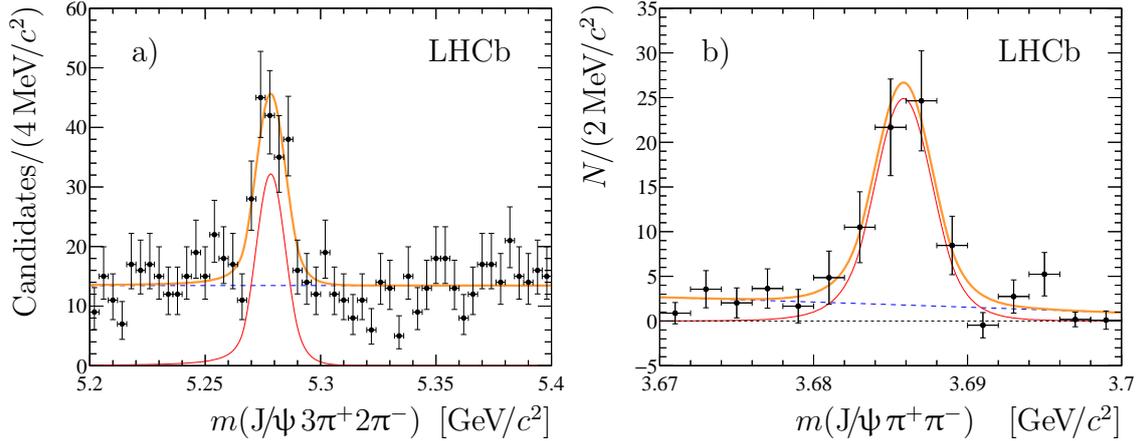
\begin{figure}[t]
  \setlength{\unitlength}{1mm}
  \centering
   \begin{picture}(150,60)
    \put( 0,0){\includegraphics*[width=75mm]{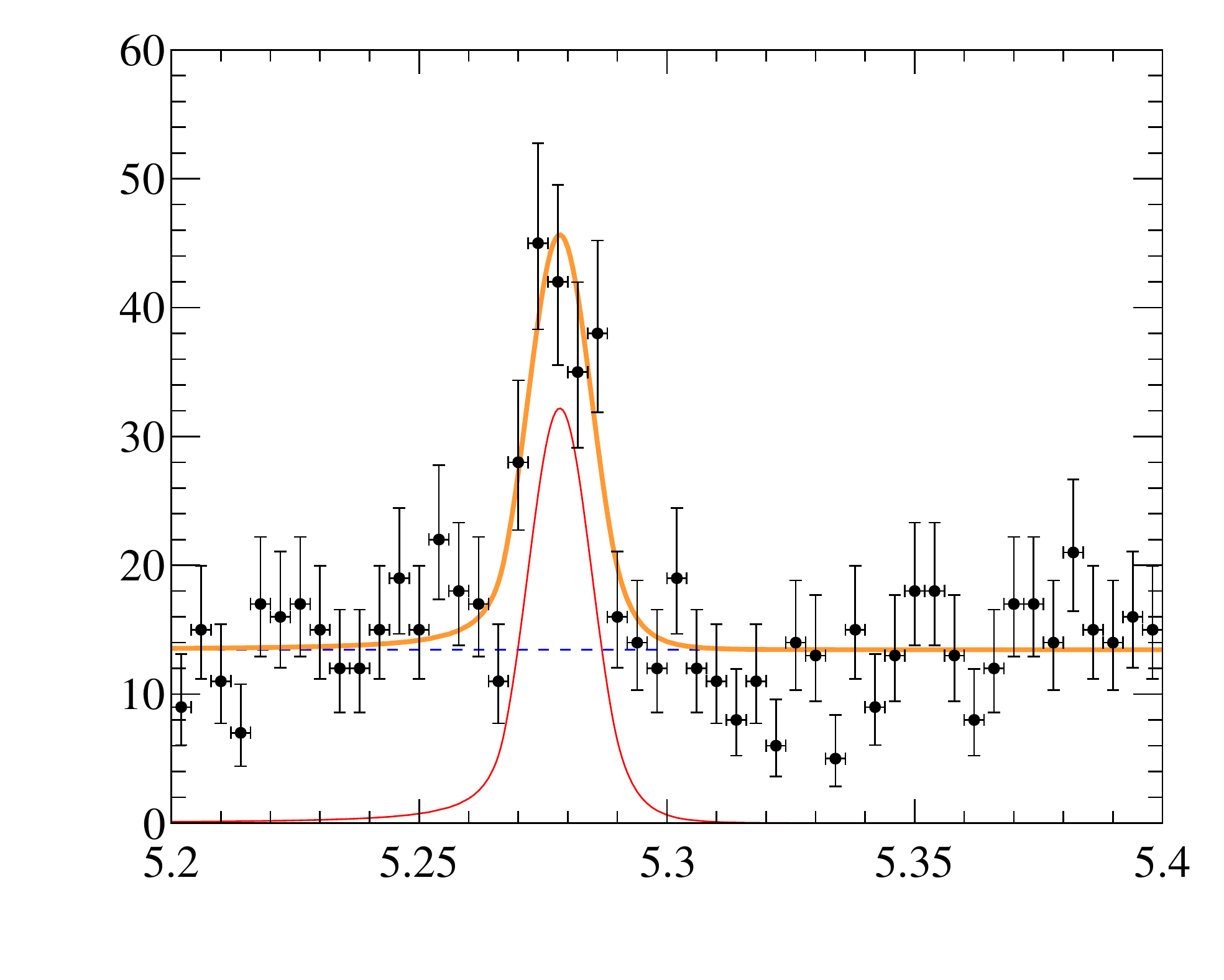}}
    \put(75,0){\includegraphics*[width=75mm]{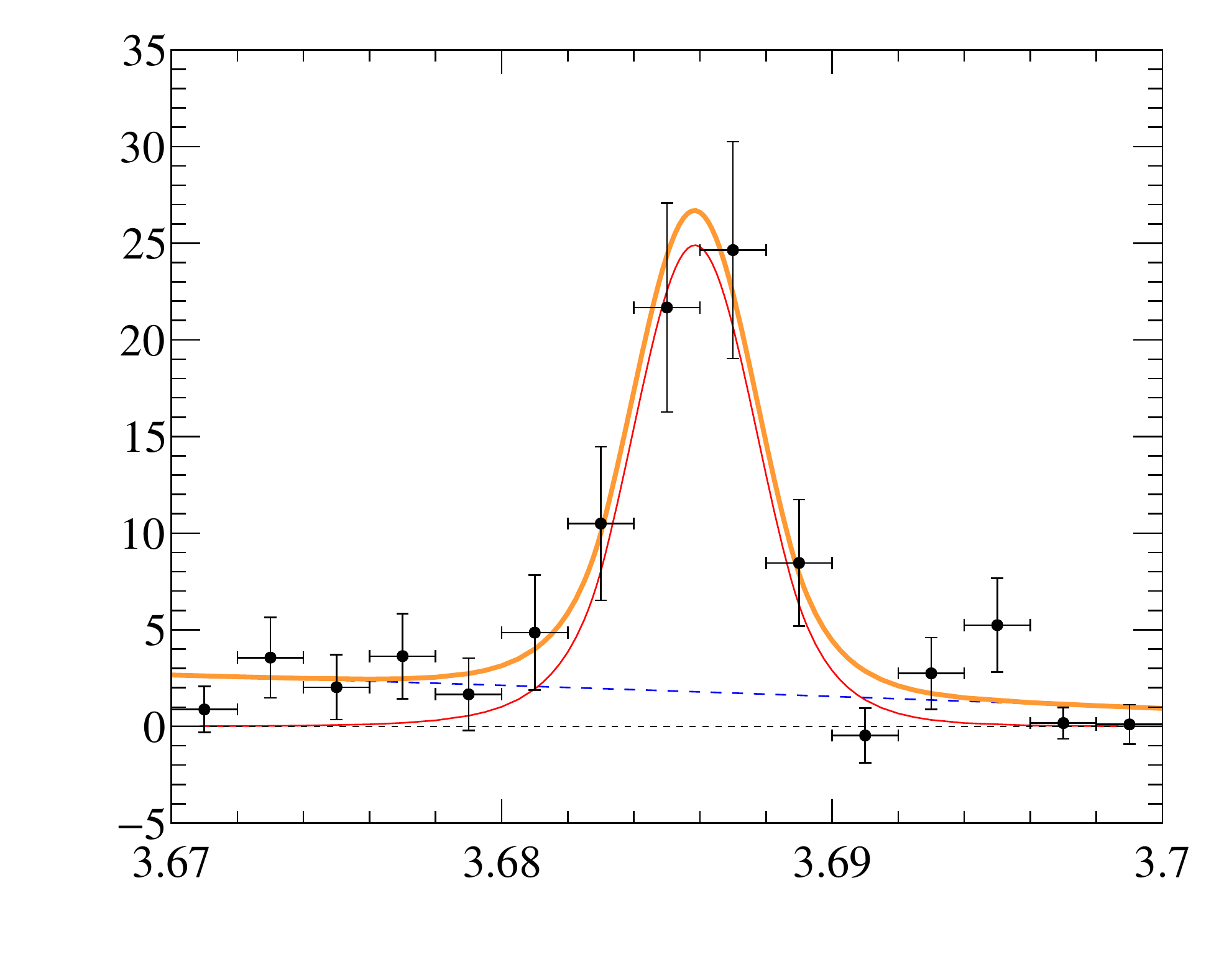}}
    \put(   0,16 )  {\begin{sideways}Candidates/$(4\mevcc)$\end{sideways}  }
    \put(  75,33 )  {\begin{sideways}$N/(2\mevcc)$\end{sideways}  }
    \put(  25,1  )  { $m(\fivepi)$}
    \put( 102,1  )  { $m(\jpsipipi)$}
    \put(  55,1  )  { $\left[\!\gevcc\right]$ }
    \put( 130,1  )  { $\left[\!\gevcc\right]$ }
    \put(  55,50  ) {LHCb}
    \put( 130,50  ) {LHCb}
    \put(  16,50  ) {a)}
    \put(  91,50  ) {b)}
  \end{picture}
  \caption { \small
    (a)~Mass distributions of the~selected \btofivepi~candidates. 
    (b)~Sum of mass distributions for all background\nobreakdash-subtracted \jpsipipi~combinations.
    The~total fit function is shown with thick solid\,(orange) lines and
    the~signal contribution with thin solid\,(red) lines.
    The~dashed\,(blue) lines represent  
    the~combinatorial background and non\nobreakdash-resonance component  
    for plots~(a) and~(b), respectively.
  }
  \label{fig:5pi}
\end{figure}

The~mass distribution for selected \btofivepi candidates
is shown in~Fig.~\ref{fig:5pi}(a). 
The~signal yield is determined with an~extended unbinned maximum likelihood fit to the~distribution. 
The~signal is modelled with a~Gaussian function with power law tails on both sides~\cite{LHCb-PAPER-2011-013}, 
where the~tail parameters are fixed from simulation 
and the~peak position and the~width of the~Gaussian function are allowed to vary.
The~combinatorial background is modelled with a~uniform distribution. 
No~peaking backgrounds from misreconstructed or 
partially reconstructed decays of beauty hadrons 
are expected in the~fit range.
The~resolution parameter obtained from the~fit is found to be~$6\pm1\mevcc$ 
and  is in good agreement with the~expectation from simulation. 
The~observed signal yield is~$139\pm18$.

The~statistical significance for the~observed signal is determined as 
\mbox{$\mathcal{S}_{\sigma}=\sqrt{-2\log \mathcal{L}_\mathrm{B}/\mathcal{L}_{\mathrm{S+B}}}$},
where ${\mathcal{L}_{\mathrm{S+B}}}$ and 
${\mathcal{L}_{\mathrm{B}}}$ denote the~likelihood associated 
with the~signal\nobreakdash-plus\nobreakdash-background and 
background\nobreakdash-only hypothesis, 
respectively. 
The~statistical significance of the~\btofivepi signal is in excess of 10~standard deviations.

For the~selected \Bp~candidates, the~existence of a resonant structure is searched for 
in the~$\jpsipipi$~combinations of final\nobreakdash-state particles. 
There are six possible \jpsipipi~combinations that can be formed 
from the~\fivepi final state.
The~background\nobreakdash-subtracted distribution of all six possible 
combinations in the~narrow range around the~known \psitwos~meson mass
is shown in Fig.~\ref{fig:5pi}(b), where each event enters six times. 
The~$sPlot$~technique is used for background subtraction~\cite{Pivk:2004ty} with the~\fivepi mass 
as the~discriminating variable.
The~signal yield of \btopsipi is determined using an~extended unbinned maximum likelihood fit
to the~background-subtracted $\jpsi\pip\pim$~mass distribution.
The~\psitwos component is modelled with a~Gaussian function with power law tails on both sides, 
where the~tail parameters are fixed from simulation. 
The~non\nobreakdash-resonant component 
is modelled with the~phase\nobreakdash-space shape multiplied by a~linear function. 
The~mass resolution obtained from the~fit is~$1.9\pm0.3\mevcc$, in good agreement 
with the expectation from simulation. 
The~observed signal yield is~$61\pm10$.

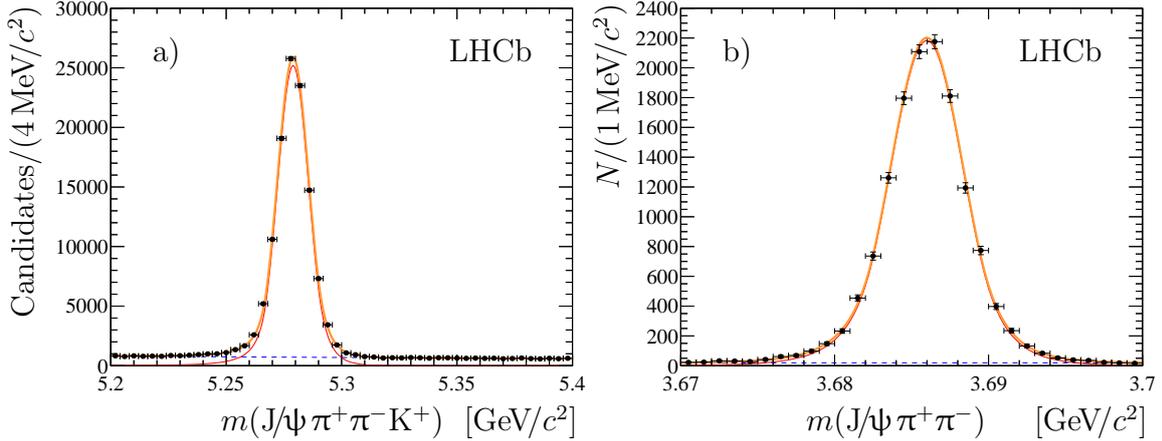
\begin{figure}[t]
  \setlength{\unitlength}{1mm}
  \centering
  \begin{picture}(150,60)
    \put( 0,0){\includegraphics*[width=75mm]{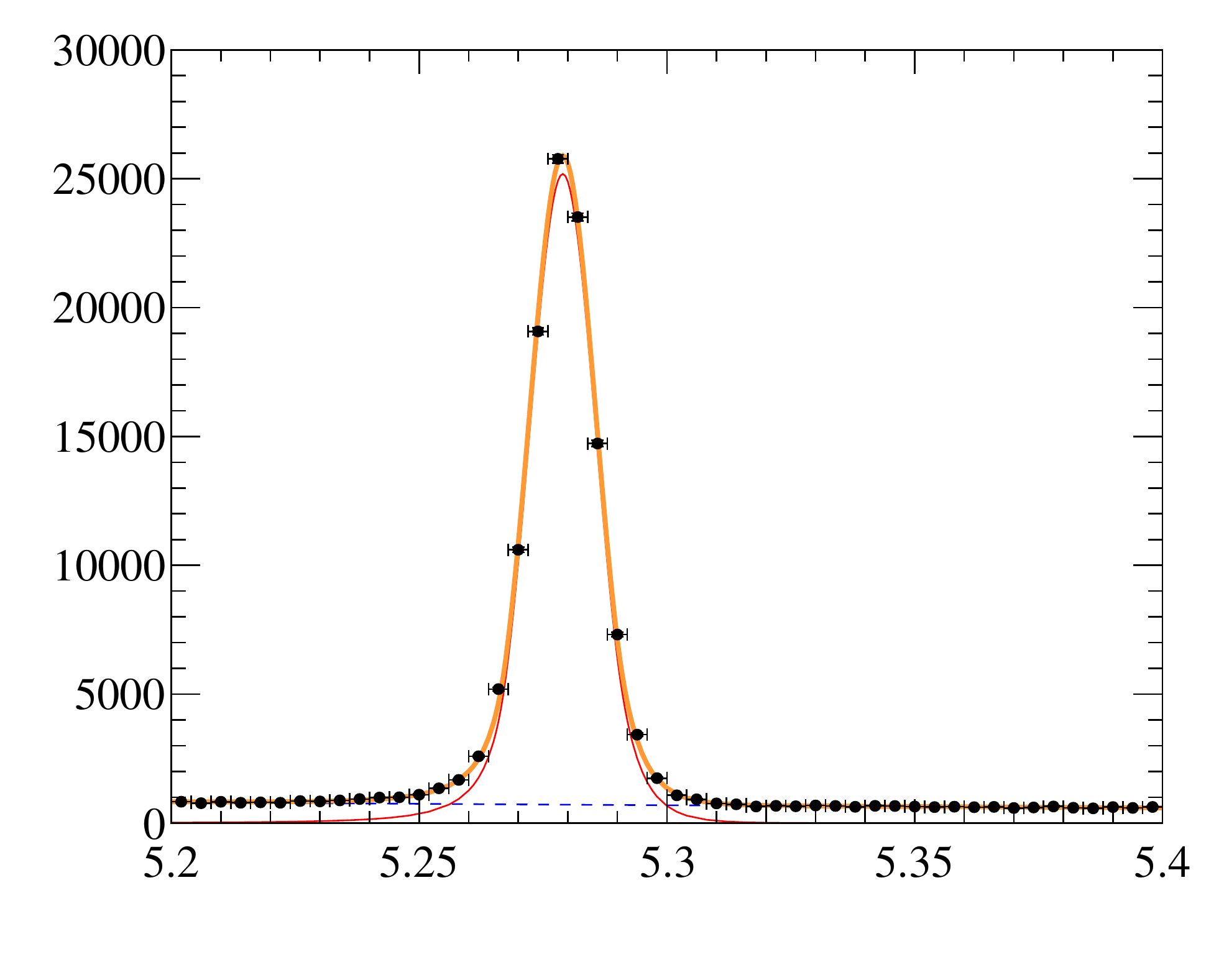}}
    \put(75,0){\includegraphics*[width=75mm]{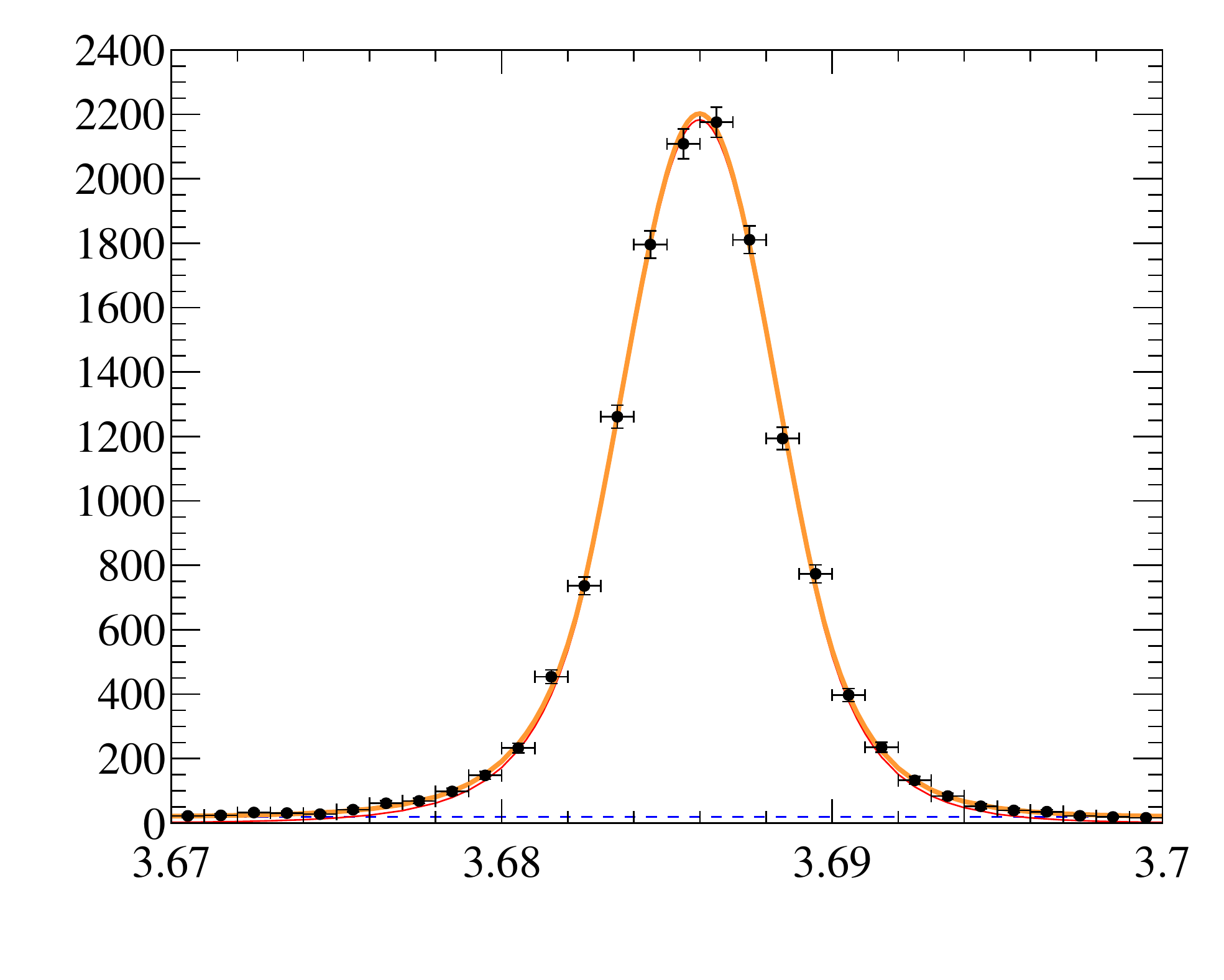}}
    \put(  -3,16 )  {\begin{sideways}Candidates/$(4\mevcc)$\end{sideways}  }
    \put(  74,33 )  {\begin{sideways}$N/(1\mevcc)$\end{sideways}  }
    \put(  25,1  )  {$m(\jpsipipi\Kp)$}
    \put( 102,1  )  {$m(\jpsipipi)$}
    \put(  57,1  )  {$\left[\!\gevcc\right]$ }
    \put( 132,1  )  {$\left[\!\gevcc\right]$ }
    \put(  55,50  ) {LHCb}
    \put( 130,50  ) {LHCb}
    \put(  16,50  ) {a)}
    \put(  91,50  ) {b)}
  \end{picture}
  \caption { \small
    Mass distributions (a) of the selected \btopsiks~candidates and 
    (b) background\nobreakdash-subtracted \jpsipipi combination.
    The~total fit function is shown with thick solid\,(orange) lines and
    the~signal contribution with thin solid\,(red) lines.
    The~dashed\,(blue) lines represent  
    the~combinatorial background and non\nobreakdash-resonance component  
    for plots~(a) and~(b), respectively.
  }
  \label{fig:psik}
\end{figure}

The~\btopsik decay 
is used as a~normalisation channel for 
the~measurements of the~relative branching fractions. 
The~mass distribution for selected \mbox{$\Bu\to\jpsi\pip\pim\Kp$} 
candidates is shown in Fig.~\ref{fig:psik}(a). 
An~extended unbinned maximum likelihood fit to the~distribution 
is performed using the~model described above for the~signal and 
an~exponential function for the~background.
The~mass resolution parameter obtained from the~fit is~$6.60\pm0.02\mevcc$, 
again in good agreement with the~expectations from simulation. 
The~background-subtracted mass distribution of the~\jpsipipi system in the~region 
of the~\psitwos mass is shown in Fig.~\ref{fig:psik}(b).

The signal yield of \btopsik is determined using an extended unbinned maximum likelihood fit 
to the~\jpsipipi distribution, 
where the background is subtracted using the $sPlot$ technique with the~$\jpsipipi\Kp$ mass 
as the~discriminating variable. 
The~\psitwos and the~non\nobreakdash-resonant components are modelled with
the~same functions used for the~signal channel.
The~mass resolution obtained from the fit is~\mbox{$2.35\pm0.02\mevcc$}. 
The~signal yields are summarized in Table~\ref{tab:fit_results}.

\begin{table}[t]
  \centering
  \caption{\small Signal yields, $N$, of \Bp decay channels. Uncertainties are statistical only.}
  \begin{tabular}{lc}
    Channel        & $N$(\Bp)               \\ \hline
    \btofivepi     & $\phantom{0}139\pm18$  \\
    \btopsipi      & $\phantom{00}61\pm10$  \\
    \btopsik       & $13554\pm118$          \\
  \end{tabular}
  \label{tab:fit_results}
\end{table}

%% file: multipion.tex
\section{Study of the multipion system}\label{seq:multi}

\begin{figure}[t]
  \setlength{\unitlength}{1mm}
  \centering
  \begin{picture}(150,60)
    \put( 0,0){\includegraphics*[width=75mm]{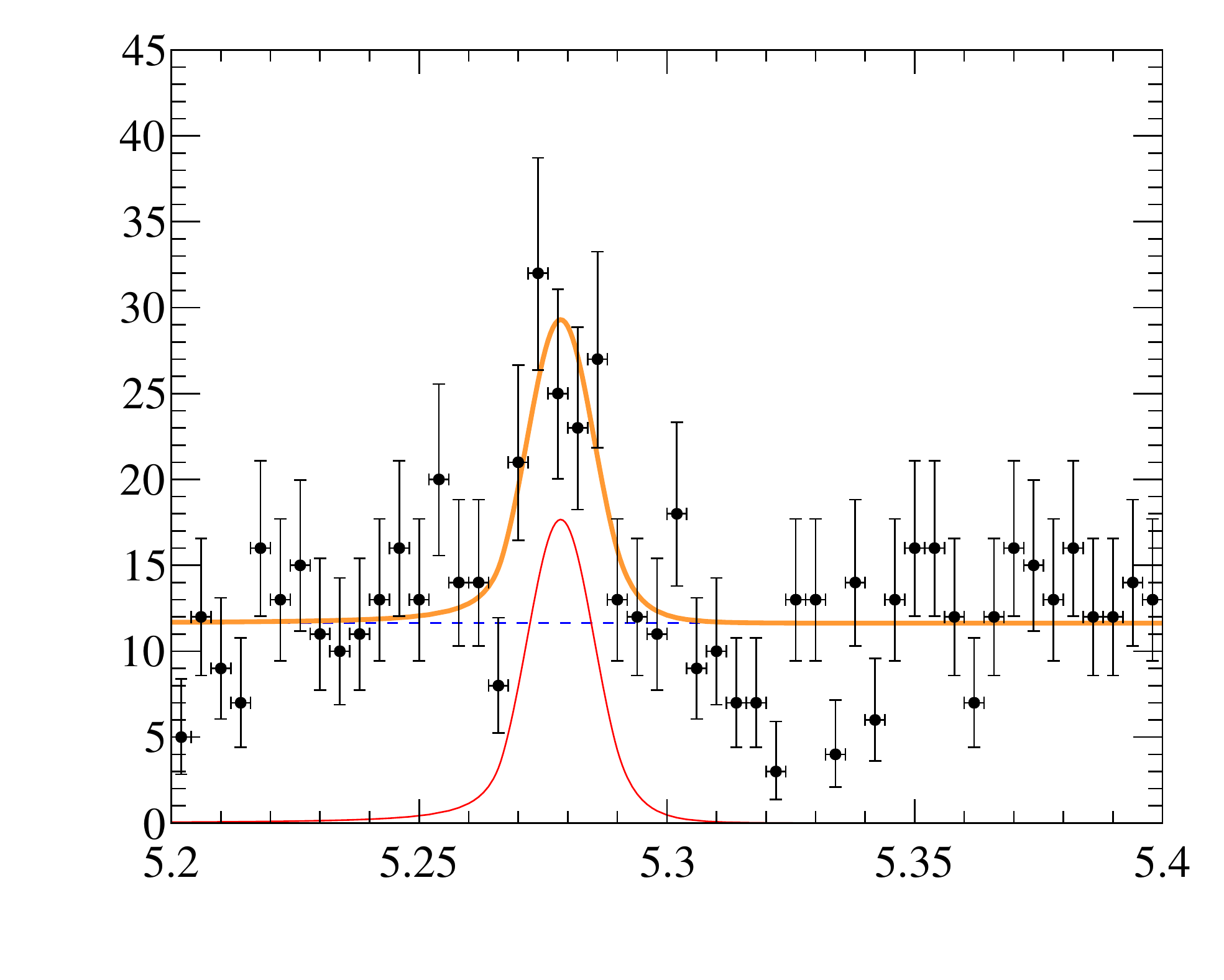}}
    \put(75,0){\includegraphics*[width=75mm]{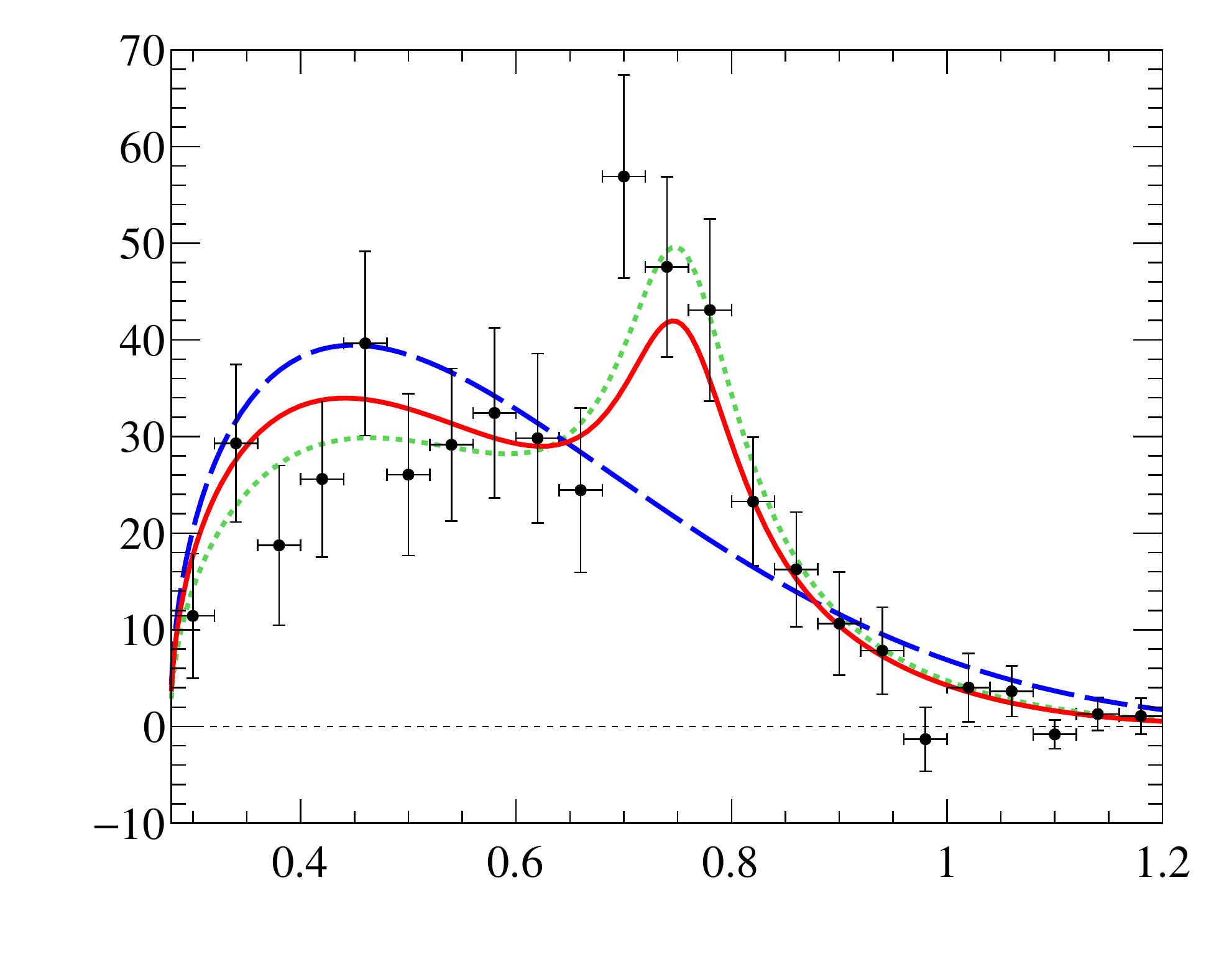}}
    \put(  -1,16 )  {\begin{sideways}Candidates/$(4\mevcc)$\end{sideways}  }
    \put(  75,31 )  {\begin{sideways}$N/(40\mevcc)$\end{sideways}  }
    \put(  25,1  )  {$m(\fivepi)$}
    \put( 102,1  )  {$m(\pip\pim)$}
    \put(  57,1  )  {$\left[\!\gevcc\right]$ }
    \put( 132,1  )  {$\left[\!\gevcc\right]$ }
    \put(  55,50  ) {LHCb}
    \put( 130,50  ) {LHCb}
    \put(  16,50  ) {a)}
    \put(  91,50  ) {b)}
  \end{picture}
  \caption { \small
    (a)~Mass distribution of the~selected \btofivepi~candidates with 
    the~additional requirement 
    of every \jpsipipi combination to be outside of $\pm6\mevcc$ around 
    the~known \psitwos~mass. 
    The~total fit function,
    the~\Bu~signal contribution and
    the~combinatorial background are shown
    with thick solid\,(orange), 
    thin solid\,(red) and dashed\,(blue) lines, respectively.
    (b)~Sum of mass distributions for all possible background\nobreakdash-subtracted 
    $\pip\pim$~combinations. 
    The~factorisation\nobreakdash-based model prediction is shown by a~solid\,(red) line, 
    and the~expectation from the~phase\nobreakdash-space model 
    is shown by a~dashed\,(blue) line. 
    The~total fit function, shown with a~dotted\,(green) line, is an~incoherent sum of 
    a~relativistic Breit\nobreakdash-Wigner function with 
    the~mean and natural width fixed to the~known $\Prho^{0}$~values 
    and a~phase\nobreakdash-space function multiplied by a~second\nobreakdash-order polynomial.}
  \label{fig:5pi_nr}
\end{figure}

A~search for intermediate light resonances is performed on 
the~set of events which do not decay through the~\psitwos~resonance. 
For this, the~additional criterion that the mass of every \jpsipipi~combination 
is outside $\pm6\mevcc$~around the~known 
\psitwos~meson mass~\cite{PDG} is applied. 
The~invariant-mass distribution for \btofivepi candidates selected 
with the~veto on the~\psitwos resonance is shown 
in Fig.~\ref{fig:5pi_nr}(a). 
A~clear peak, corresponding to the~non\nobreakdash-resonant decay 
$\Bu\to\jpsi3\pip2\pim$~decay is visible.
The~signal yield for this channel is determined using 
an~extended unbinned maximum likelihood fit using the~function described above.
The~observed signal yield is~$80\pm15$
with a~statistical significance of 6.8~standard deviations.

\begin{figure}[t]
  \setlength{\unitlength}{1mm}
  \centering
\begin{picture}(150,180) 
  \put( 5,135){\includegraphics*[width=60mm,height=45mm]{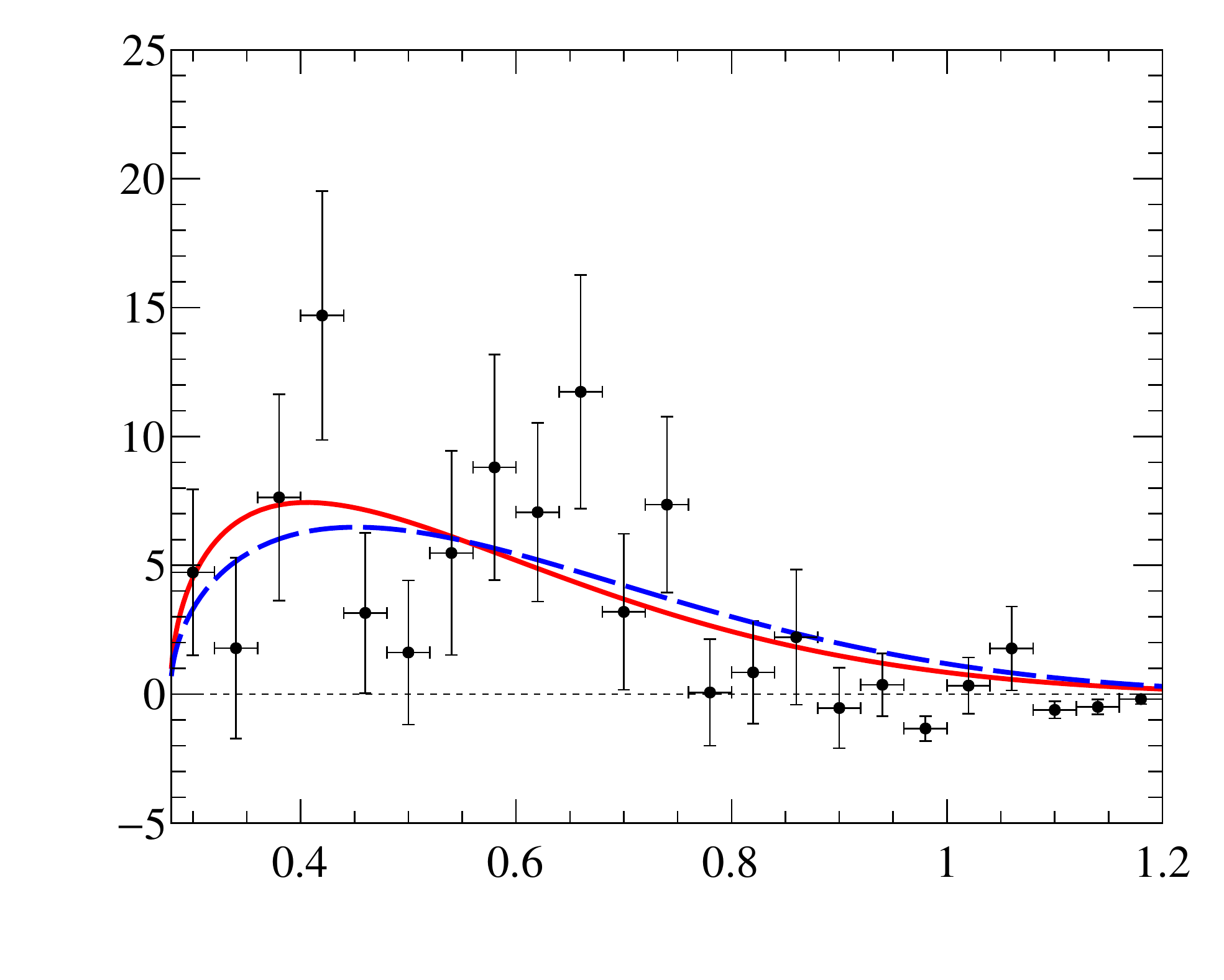}}
  \put(75,135){\includegraphics*[width=60mm,height=45mm]{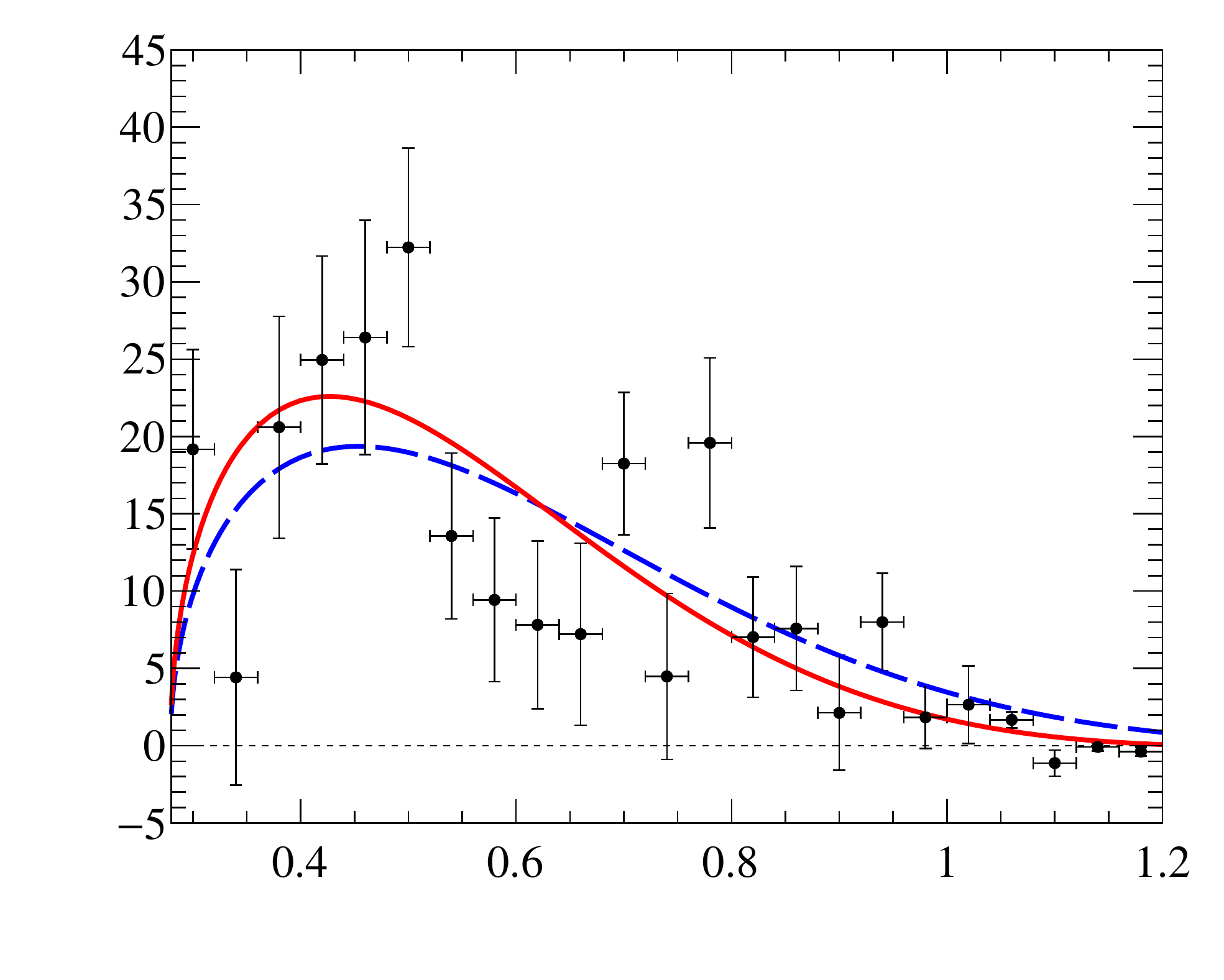}}
  \put( 5,90 ){\includegraphics*[width=60mm,height=45mm]{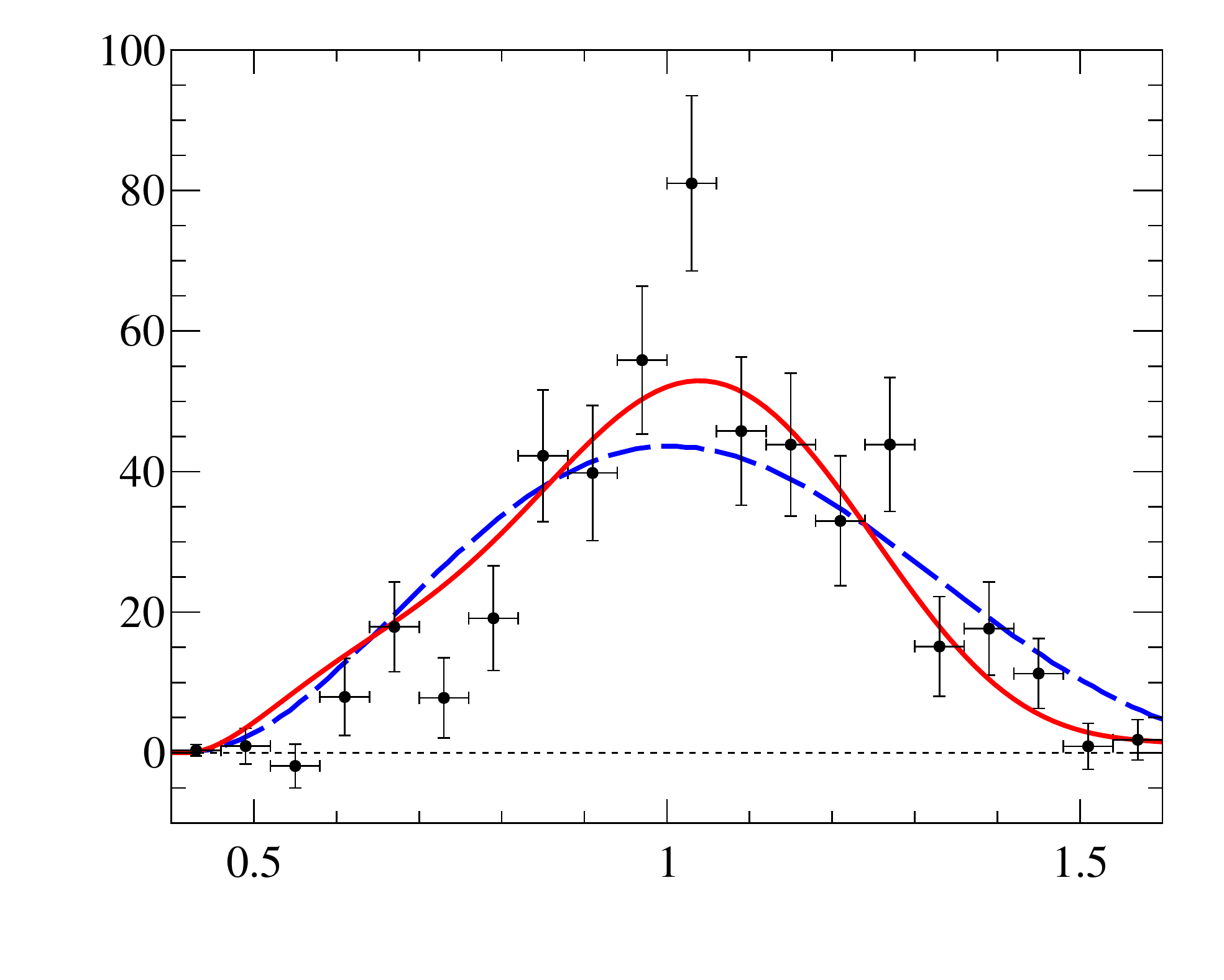}}
  \put(75,90 ){\includegraphics*[width=60mm,height=45mm]{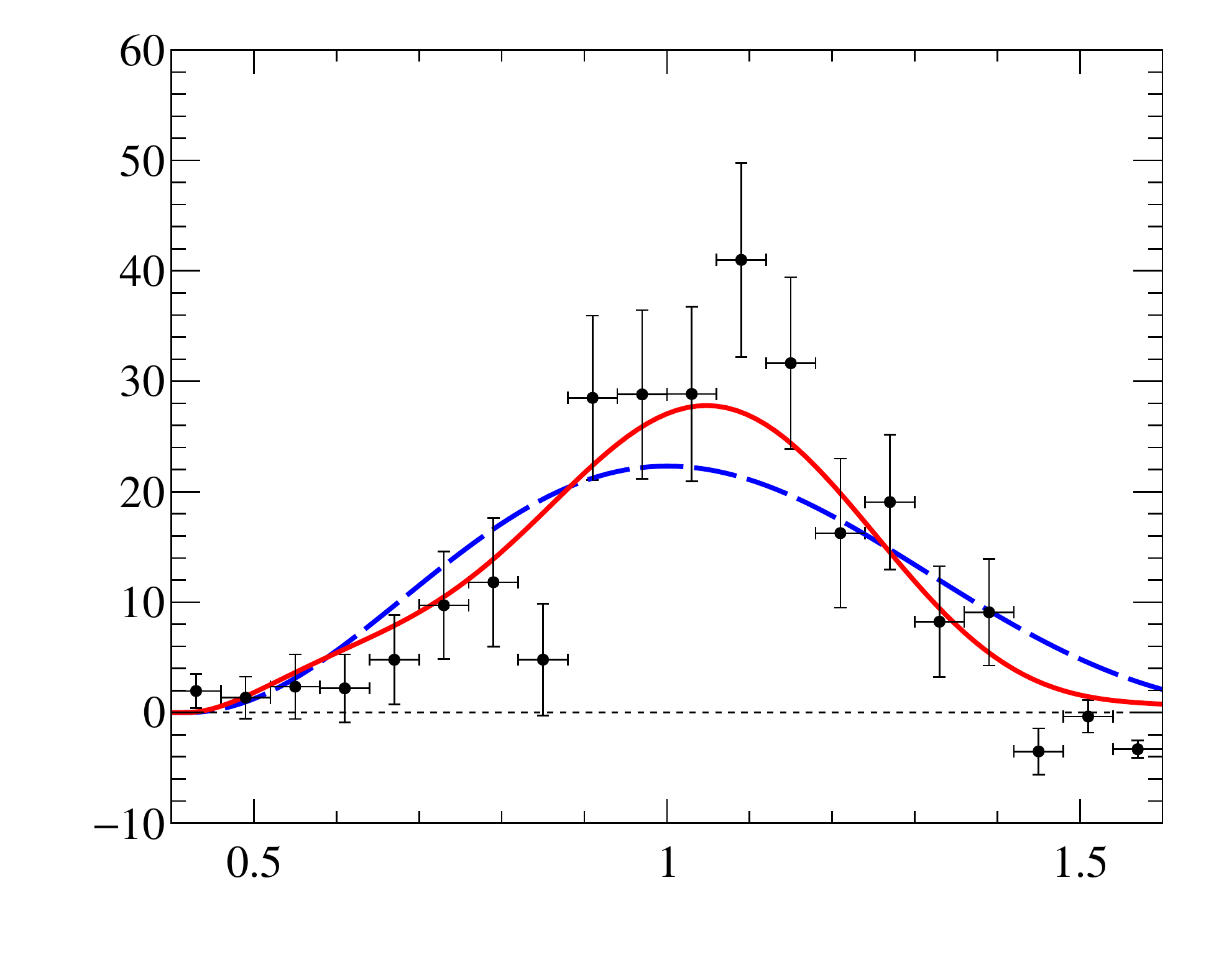}}
  \put( 5,45 ){\includegraphics*[width=60mm,height=45mm]{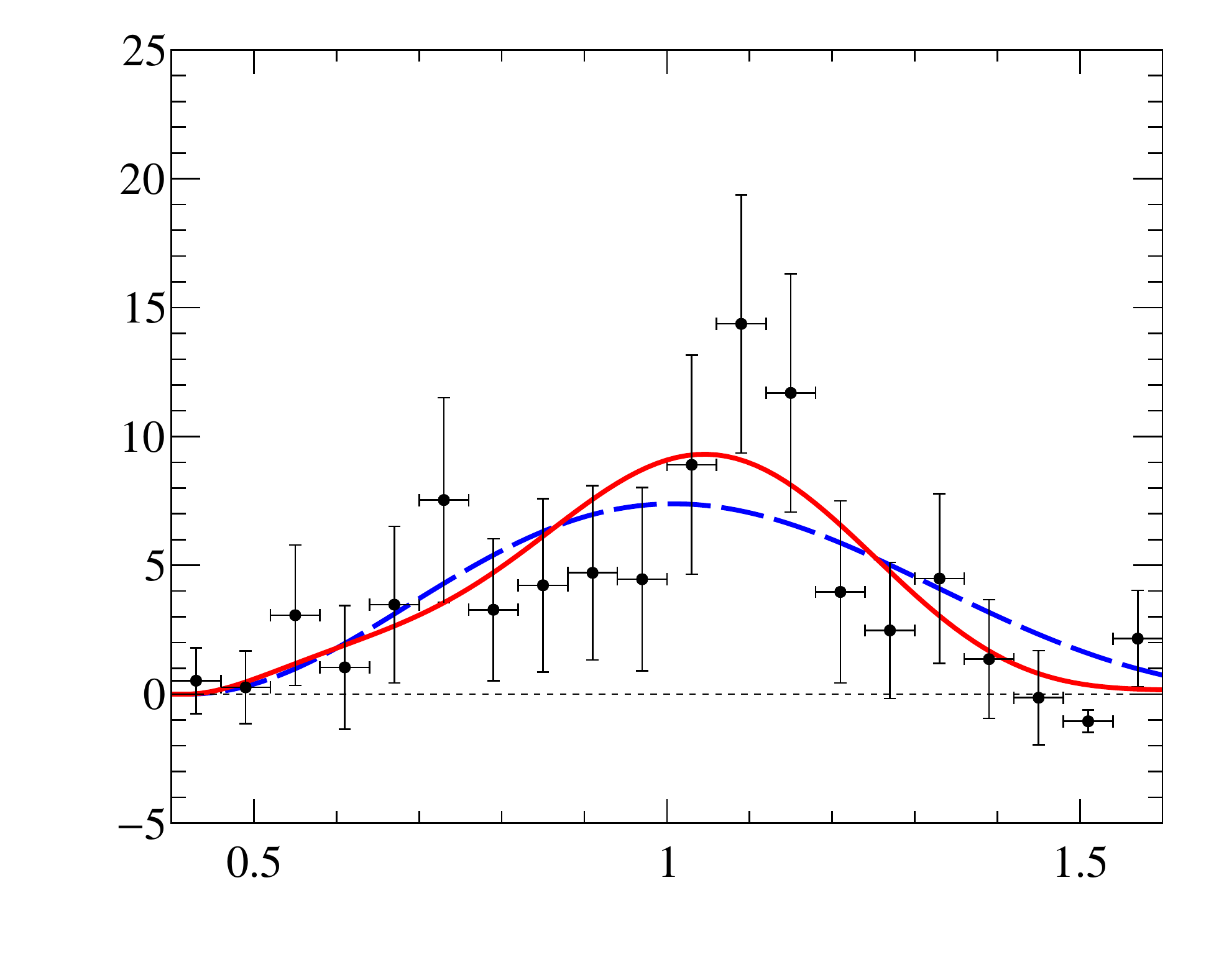}}
  \put(75,45 ){\includegraphics*[width=60mm,height=45mm]{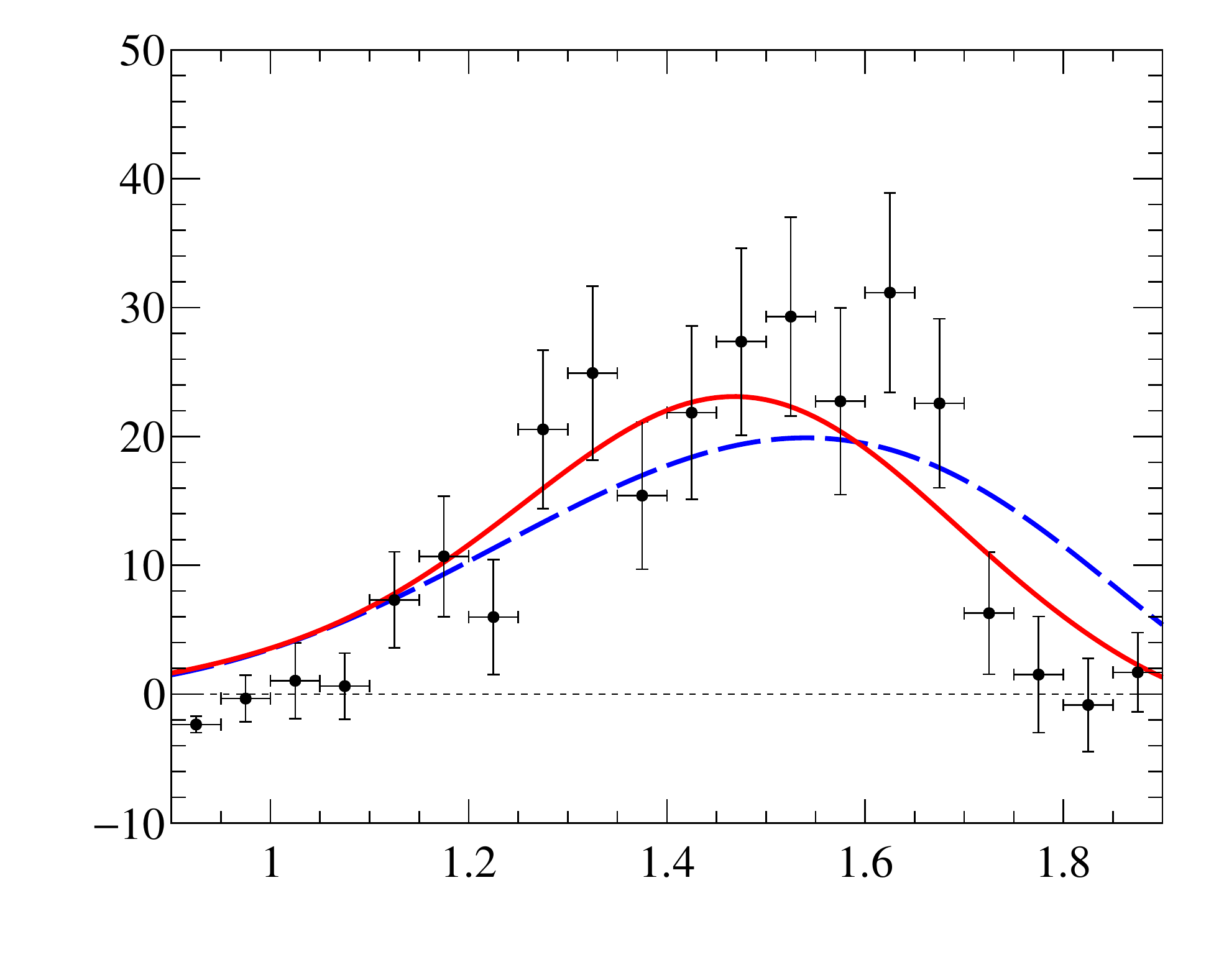}}
  \put( 5,0  ){\includegraphics*[width=60mm,height=45mm]{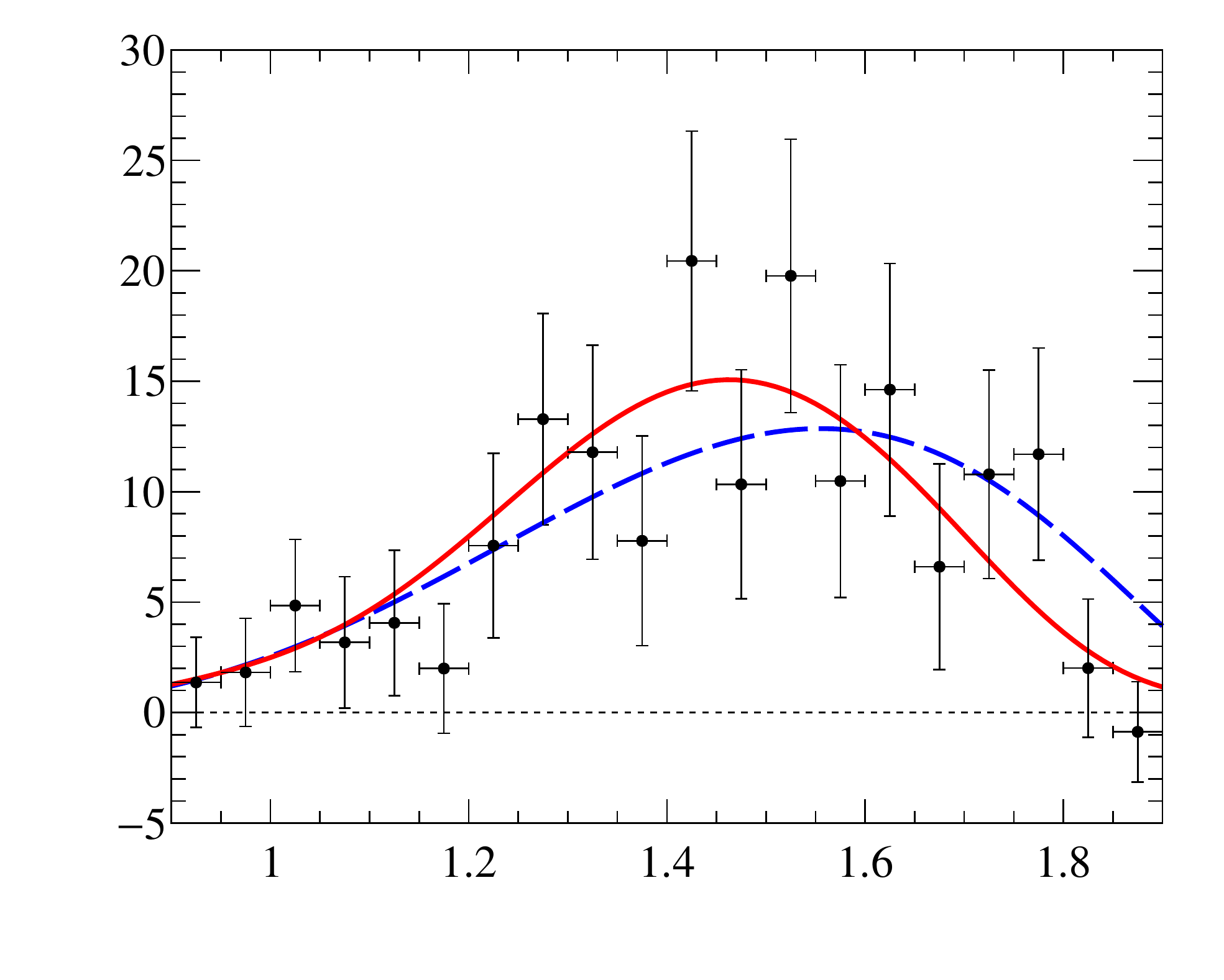}}
  \put(75,0  ){\includegraphics*[width=60mm,height=45mm]{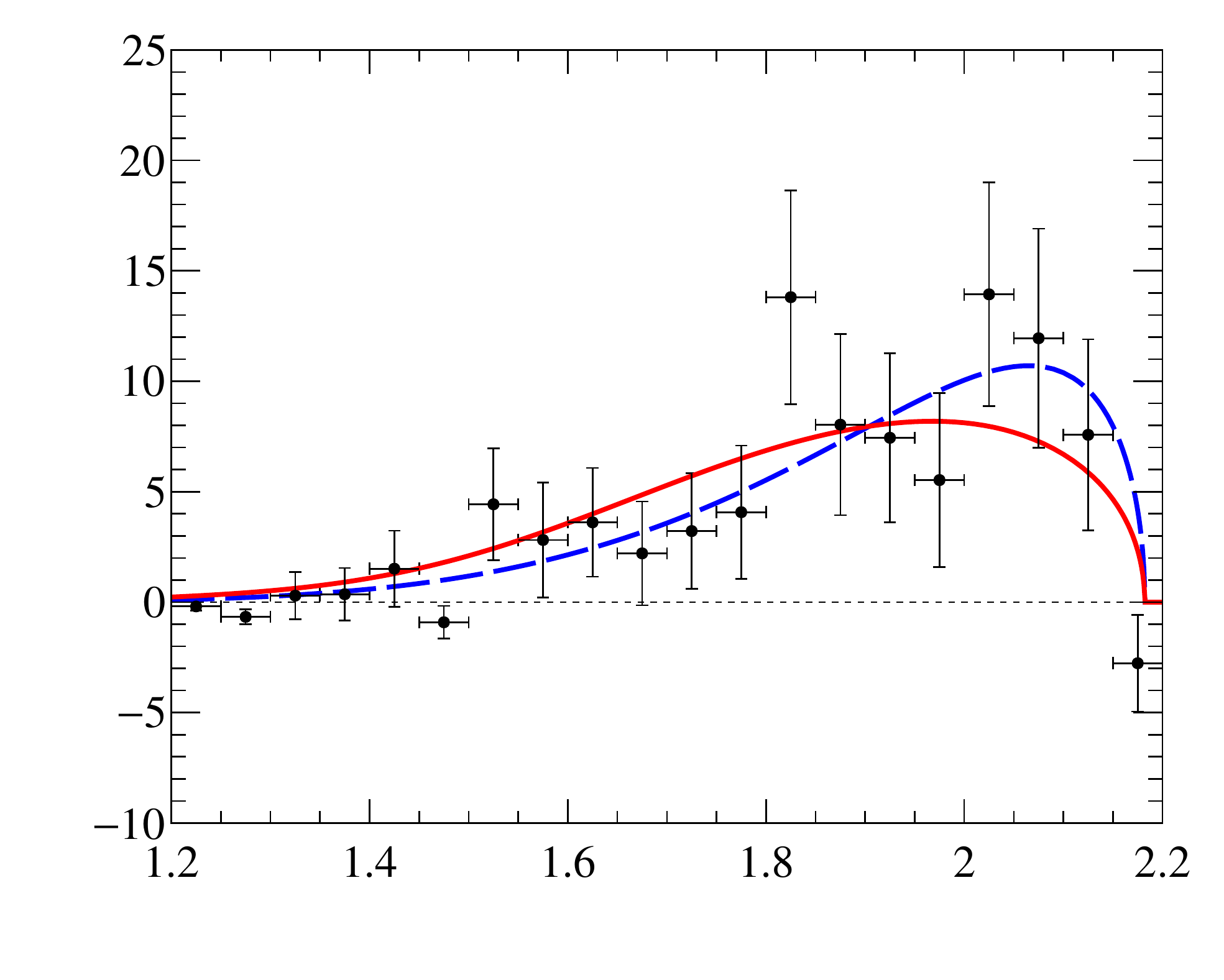}}
  \put( 16,172){\small{a)}}
  \put( 86,172){\small{b)}}
  \put( 16,127){\small{c)}}
  \put( 86,127){\small{d)}}
  \put( 16,82 ){\small{e)}}
  \put( 86,82 ){\small{f)}}
  \put( 16,37 ){\small{g)}}
  \put( 86,37 ){\small{h)}}
  \put( 49,172){\small{LHCb}}
  \put(119,172){\small{LHCb}}
  \put( 49,127){\small{LHCb}}
  \put(119,127){\small{LHCb}}
  \put( 49,82 ){\small{LHCb}}
  \put(119,82 ){\small{LHCb}}
  \put( 49,37 ){\small{LHCb}}
  \put(119,37 ){\small{LHCb}}
  \put( 5,153){\begin{sideways}\small{$N/(40\mevcc)$}\end{sideways}}
  \put(75,153){\begin{sideways}\small{$N/(40\mevcc)$}\end{sideways}}
  \put( 5,108){\begin{sideways}\small{$N/(60\mevcc)$}\end{sideways}}
  \put(75,108){\begin{sideways}\small{$N/(60\mevcc)$}\end{sideways}}
  \put( 5,63 ){\begin{sideways}\small{$N/(60\mevcc)$}\end{sideways}}
  \put(75,63 ){\begin{sideways}\small{$N/(50\mevcc)$}\end{sideways}}
  \put( 5,18 ){\begin{sideways}\small{$N/(50\mevcc)$}\end{sideways}}
  \put(75,18 ){\begin{sideways}\small{$N/(50\mevcc)$}\end{sideways}}
  \put( 26,135){\small{$m(\pim\pim)$}}
  \put( 96,135){\small{$m(\pip\pip)$}}
  \put( 26,90 ){\small{$m(\pip\pip\pim)$}}
  \put( 96,90 ){\small{$m(\pip\pim\pim)$}}
  \put( 26,45 ){\small{$m(\pip\pip\pip)$}}
  \put( 96,45 ){\small{$m(2\pip2\pim)$}}
  \put( 26,0  ){\small{$m(3\pip\pim)$}}
  \put( 96,0  ){\small{$m(3\pip2\pim)$}}
  \put( 48,135){\small{$\left[\!\gevcc\right]$}}
  \put(118,135){\small{$\left[\!\gevcc\right]$}}
  \put( 48,90 ){\small{$\left[\!\gevcc\right]$}}
  \put(118,90 ){\small{$\left[\!\gevcc\right]$}}
  \put( 48,45 ){\small{$\left[\!\gevcc\right]$}}
  \put(118,45 ){\small{$\left[\!\gevcc\right]$}}
  \put( 48,0  ){\small{$\left[\!\gevcc\right]$}}
  \put(118,0  ){\small{$\left[\!\gevcc\right]$}}
  \end{picture}
  \caption {\small 
    Distributions of 
    (a)\,$\pim\pim$, 
    (b)\,$\pip\pip$, 
    (c)\,$\pip\pip\pim$, 
    (d)\,$\pip\pim\pim$, 
    (e)\,$\pip\pip\pip$, 
    (f)\,$2\pip2\pim$, 
    (g)\,$3\pip\pim$ and 
    (h)\,$3\pip2\pim$~masses in the~\btofivepi decay. 
    The~prediction from the factorisation\nobreakdash-based model is shown by solid\,(red) lines, 
    and the~expectation from the~phase\nobreakdash-space model is shown by dashed\,(blue) lines.
  }
  \label{fig:pions_nr}
\end{figure}
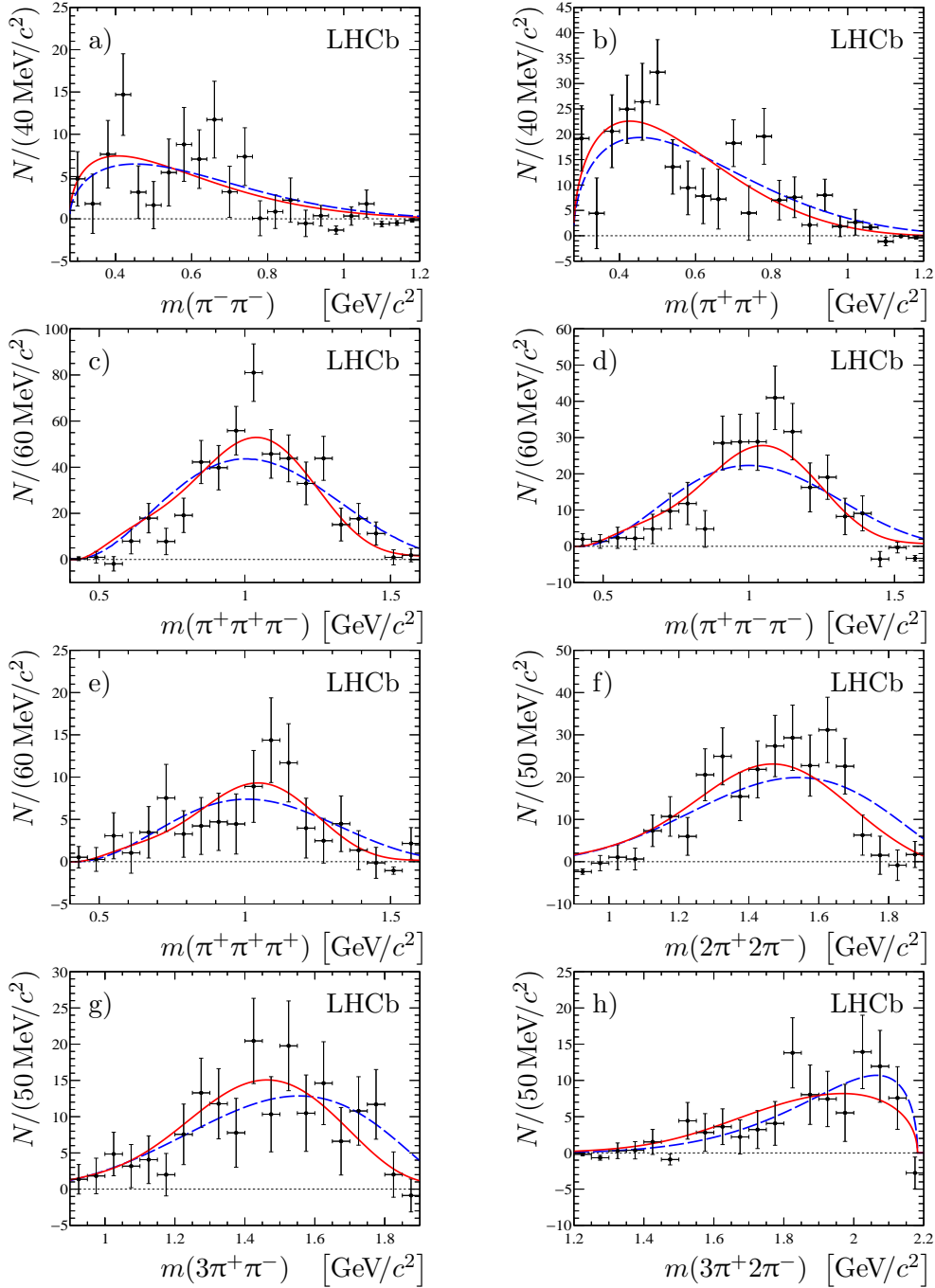

The~resonance structure is investigated 
in the~$\pip\pim$, 
$\pip\pip$, 
$\pim\pim$, 
$\pip\pip\pim$,
$\pip\pim\pim$,
$\pip\pip\pip$,
$2\pip2\pim$,
$3\pip\pim$ and~$3\pip2\pim$~combinations
of final\nobreakdash-state particles using the~$sPlot$ technique, 
with the~reconstructed $\fivepi$~mass as the~discriminating variable. 
The~resulting background\nobreakdash-subtracted mass distribution 
of all possible $\pip\pim$~combinations is shown in Fig.~\ref{fig:5pi_nr}(b), 
along with the~theoretical predictions from the~factorisation approach 
and the~phase\nobreakdash-space model~\cite{Lesha,Likhoded:2013iua,Berezhnoy:2011is,Likhoded:2009ib}. 
A~structure is seen that can be associated to the~$\Prho^{0}$~meson. 
The~distribution is fitted with a~sum of a~relativistic Breit\nobreakdash-Wigner~function 
with the~mean and natural width fixed to the~known $\Prho^{0}$~values plus 
a~phase\nobreakdash-space shape multiplied by 
a~second\nobreakdash-order polynomial. 
No~significant narrow structures are observed for other 
\mbox{multipion} combinations. 
The~distributions for all other combinations of pions 
are compared with predictions of both 
a~factorisation approach and a~phase\nobreakdash-space model,
as shown in Fig.~\ref{fig:pions_nr}. 
For all fits the \chisq per degree of freedom, $\chisq/\mathrm{ndf}$, is given in Table~\ref{tab:chisq_nr}.
The~prediction from the~factorisation approach 
is found to be in somewhat better agreement 
with the~data than that from the~phase\nobreakdash-space model, 
giving better $\chisq/\mathrm{ndf}$~values for eight out of 
nine distributions examined.

\begin{table}[t]
  \centering
  \caption{ \small
    The~\chisq per degree of freedom for the 
    factorisation\nobreakdash-based 
    and phase\nobreakdash-space models
    for the multipion system in non\nobreakdash-resonant 
    \mbox{$\Bu\to\jpsi 3\pip2\pim$}~decays.
  } \label{tab:chisq_nr}
  \vspace*{3mm}
  \begin{tabular*}{0.85\textwidth}{@{\hspace{5mm}}l@{\extracolsep{\fill}}cc@{\hspace{5mm}}}
    Multipion system   &    Factorisation model  &  Phase-space model  \\
    \hline 
    $\pip\pim$     &   0.7 & 2.6 \\ 
    $\pim\pim$     &   2.8 & 3.7 \\ 
    $\pip\pip$     &   1.7 & 4.2 \\
    $\pip\pip\pim$ &   1.8 & 2.3 \\
    $\pip\pim\pim$ &   2.8 & 5.0 \\
    $\pip\pip\pip$ &   1.0 & 2.5 \\
    $2\pip2\pim$   &   3.5 & 4.4 \\
    $2\pip\pim$    &   0.7 & 1.0 \\
    $3\pip2\pim$   &   2.2 & 1.7 
  \end{tabular*}
\end{table}

\begin{figure}[t]
\setlength{\unitlength}{1mm}
\centering
\begin{picture}(150,120)
  \put( 0,55 ){\includegraphics*[width=70mm]{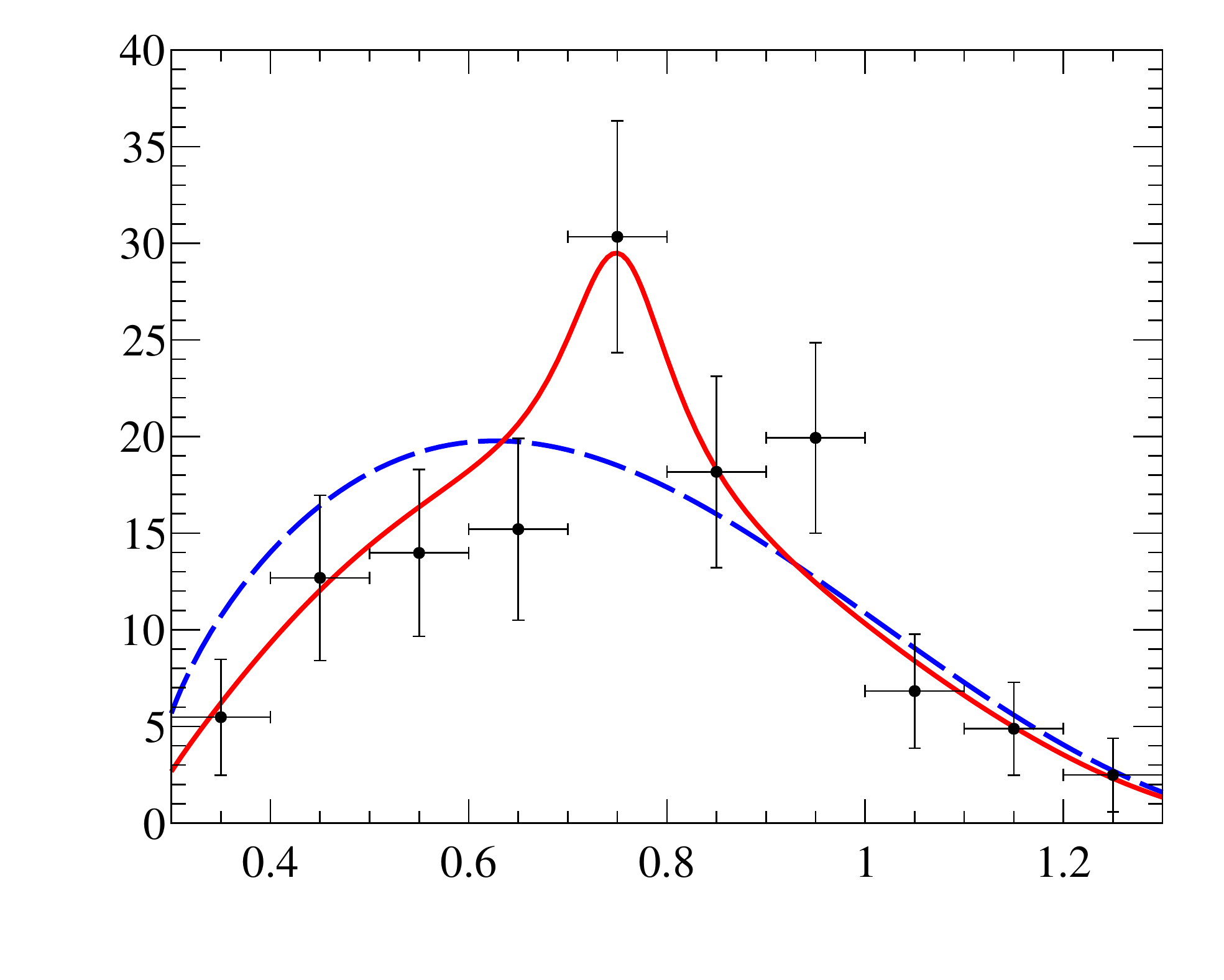}}
  \put(70,55 ){\includegraphics*[width=70mm]{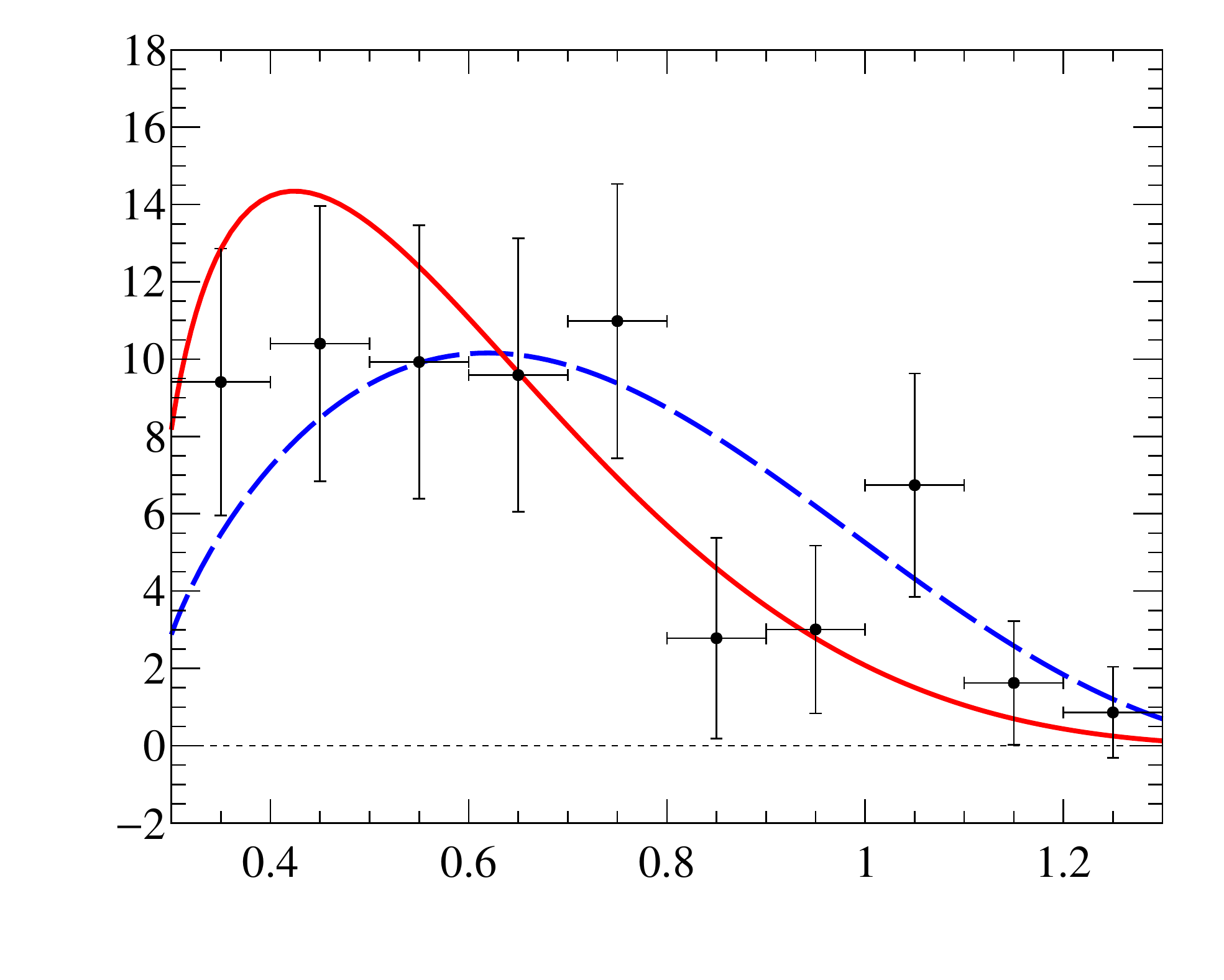}}
  \put(35,0  ){\includegraphics*[width=70mm]{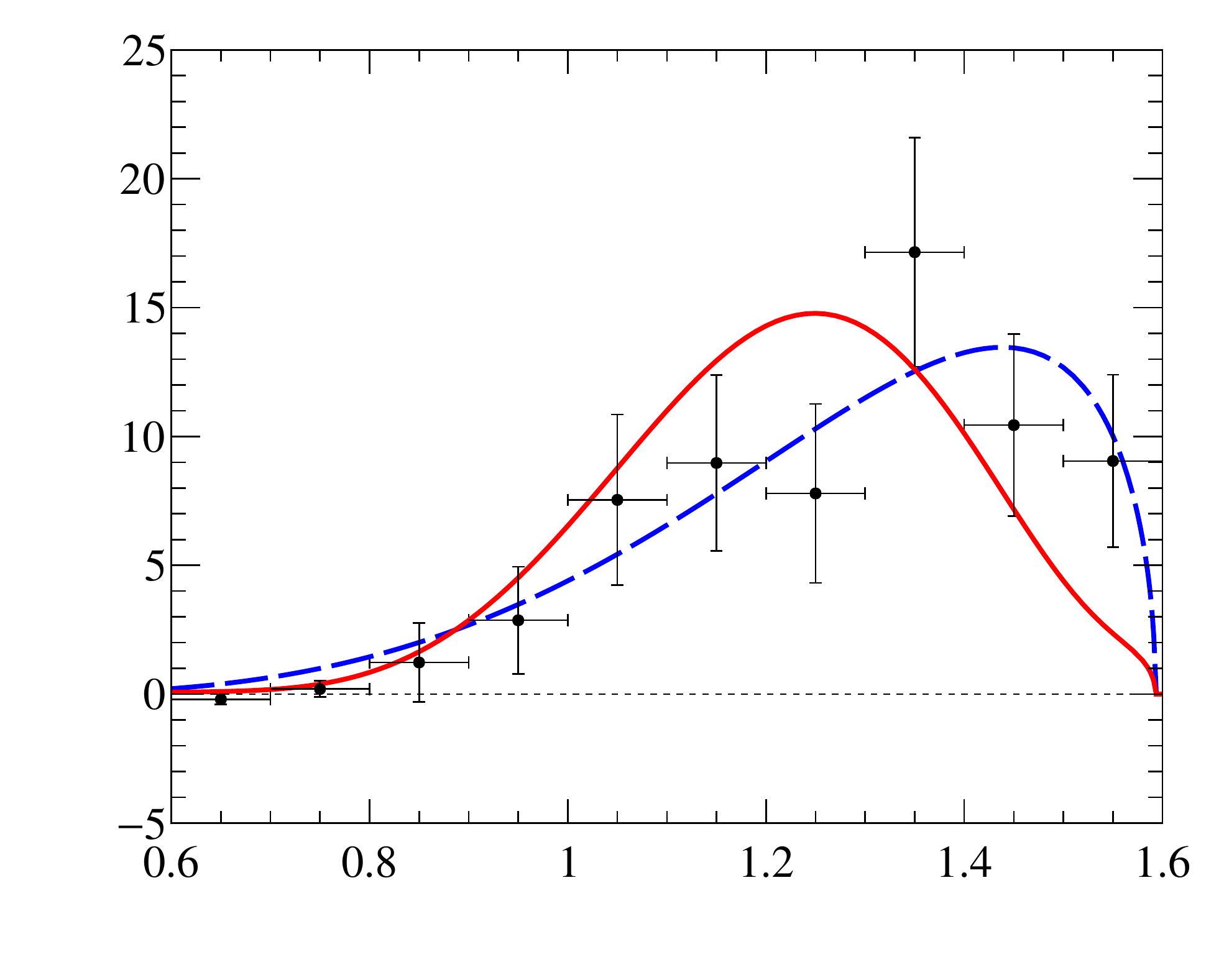}}
  \put(17,101){a)}
  \put(87,101){b)}
  \put(92,45 ){c)}
  \put( 50,101){LHCb}
  \put(120,101){LHCb}
  \put( 52,45 ){LHCb}
  \put( 0,81 ){\begin{sideways}\small{ N/(100$\mevcc$) }\end{sideways}}
  \put(71,81 ){\begin{sideways}\small{ N/(100$\mevcc$) }\end{sideways}}
  \put(35,26 ){\begin{sideways}\small{ N/(100$\mevcc$) }\end{sideways}}
  \put( 30,57 ){\small{$m(\pip\pim)$}}
  \put(100,57 ){\small{$m(\pip\pip)$}}
  \put( 65,2  ){\small{$m(\pip\pip\pim)$}}
  \put( 53,56 ){\small{$\left[\!\gevcc\right]$}}
  \put(123,56 ){\small{$\left[\!\gevcc\right]$}}
  \put( 88,1  ){\small{$\left[\!\gevcc\right]$}}
  \end{picture}
  \caption {\small 
    Distributions of 
    (a)\,$\pip\pim$, 
    (b)\,$\pip\pip$ and  
    (c)\,$\pip\pip\pim$~masses 
    in the~\mbox{$\Bp\to\psitwos\pip\pip\pim$}~decay. 
    The~predictions from the factorisation\nobreakdash-based model is shown by solid\,(red) lines, 
    and the~expectation from the~phase\nobreakdash-space model is shown by dashed\,(blue) lines. }
  \label{fig:pions_res}
\end{figure}
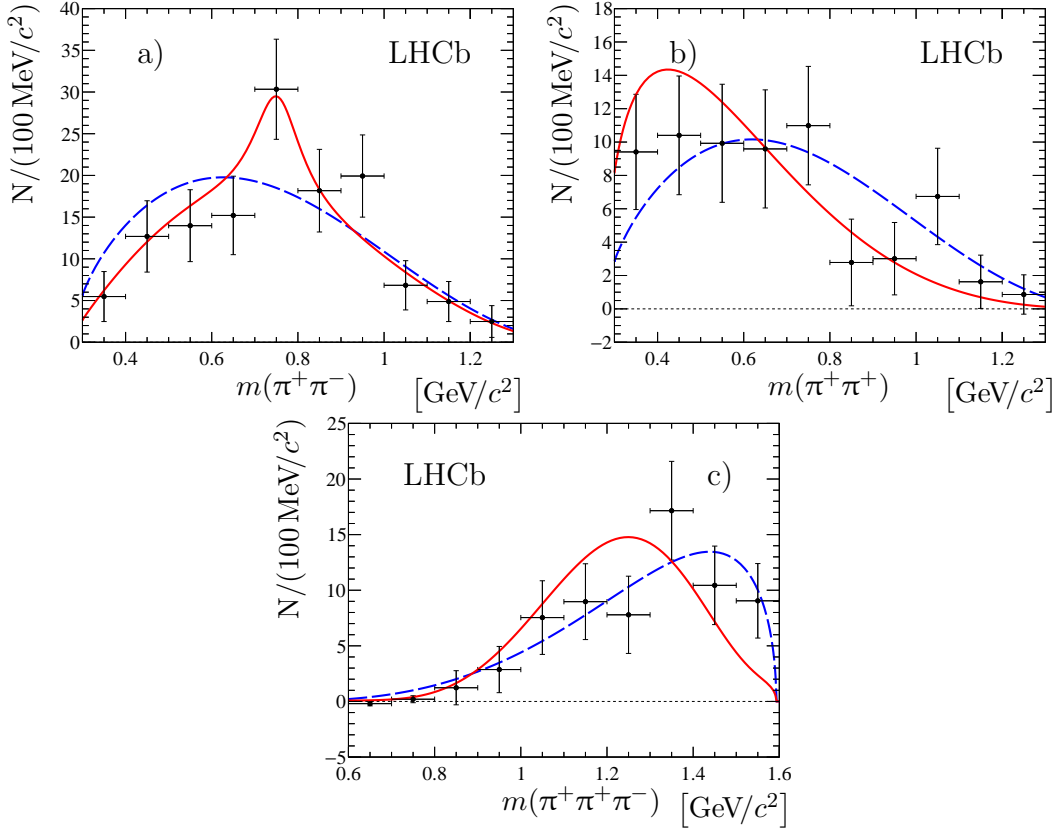

In a~similar way intermediate light resonances 
are searched for in the three\nobreakdash-pion system recoiling against
$\psitwos\to\jpsi\pip\pim$ in $\Bp\to\psitwos\pip\pip\pim$~decays.
The~resonant structure is investigated 
in the~$\pip\pim$, 
$\pip\pip$ and $\pip\pip\pim$~combinations.
The~distributions for these combinations of pions 
are compared with predictions of both 
the~factorisation approach and a~phase\nobreakdash-space model,
as shown in Fig.~\ref{fig:pions_res}. 
The corresponding $\chisq/\mathrm{ndf}$ values are summarized in Table~\ref{tab:chisq_res}.
Similarly to the~non-resonant case, the~prediction from the~factorisation approach 
is found to be in somewhat better agreement 
with the~data than that from the~phase\nobreakdash-space model. 

\begin{table}[t]
  \centering
  \caption{ \small
    The~\chisq per degree of freedom for the 
    factorisation\nobreakdash-based  
    and phase\nobreakdash-space models
    for the multipion system recoiling against $\psitwos$
    in \mbox{$\Bu\to\psitwos \pip\pip\pim$}~decays.
  } 
  \label{tab:chisq_res}
  \vspace*{3mm}
  \begin{tabular*}{0.85\textwidth}{@{\hspace{5mm}}l@{\extracolsep{\fill}}cc@{\hspace{5mm}}}
    Multipion system   &    Factorisation model  &  Phase-space model  \\
    \hline 
    $\pip\pim$     &   0.5 & 1.3 \\ 
    $\pip\pip$     &   0.8 & 0.7 \\
    $\pip\pip\pim$ &   1.3 & 1.6 
  \end{tabular*}
\end{table}

%% file: effic.tex
\section{Efficiencies and systematic uncertainties}
\label{sec:effic}

The two ratios of branching fractions defined in Eq.~\ref{eq:rate_fivepi} are measured as
\begin{align*}
  R_{5\pi} & = \dfrac{N_{\jpsi 3\pip2\pim}}{N_{\psitwos[\to\jpsi\pip\pim]\Kp}}
  \times
  \dfrac{\upvarepsilon_{\psitwos[\to\jpsi\pip\pim]\Kp}}{\upvarepsilon_{\jpsi 3\pip2\pim}}\times\BR(\psitwos\to\jpsi\pip\pim)~,
  \\
  R_{\psitwos} &= \dfrac{N_{\psitwos[\to\jpsi\pip\pim]\pip\pip\pim}}{N_{\psitwos[\to\jpsi\pip\pim]\Kp}}
  \times\dfrac{\upvarepsilon_{\psitwos[\to\jpsi\pip\pim]\Kp}}{\upvarepsilon_{\psitwos[\to\jpsi\pip\pim]\pip\pip\pim}}~,
\end{align*}
where $N_{X}$ represents the~observed signal yield and $\upvarepsilon_{X}$ denotes 
the~efficiency for the~corresponding decay. 
The~known value of~$(34.46\,\pm\,0.30)\%$~\cite{PDG} is used for the~\psitwostopsi branching fraction.

The~efficiency is determined as the~product of the~geometric acceptance 
and the~detection, reconstruction, selection and trigger efficiencies. 
The~efficiencies for hadron identification as a~function of the~kinematic parameters 
and event multiplicity are determined from data, using calibration samples of kaons and pions from the 
self-identifying decays $\Dstarp\to\Dz\pip$~followed by $\Dz\to\Km\pip$~\cite{LHCb-DP-2012-003}.
The~remaining efficiencies are determined using simulated events. 

To determine the~overall efficiency for the~\btofivepi channel, 
the~individual efficiencies for the resonant and non-resonant components are averaged 
according to the~measured proportions found in the data, 
\begin{equation*}
  k \equiv \dfrac{N_{\psitwos[\to\jpsi\pip\pim]\pip\pip\pim}}{N_{\jpsi 3\pip2\pim}} = 0.44\pm0.06~.
\end{equation*}
The~ratio $k$~is calculated taking into account the~correlation in the~observed values in the~numerator and denominator. 
The~ratios of the~efficiency for the~normalization channel $\upvarepsilon_{\psitwos\Kp}$
to the~efficiencies for resonant, $\upvarepsilon_{\psitwos \pip\pip\pim}$,
and non-resonant decays $\upvarepsilon_{\jpsi 3\pip2\pim, \mathrm{NR}}$, 
are determined to be
\begin{eqnarray*}
  \dfrac{\upvarepsilon_{\psitwos\Kp}}{\upvarepsilon_{\psitwos \pip\pip\pim}}             &= & 6.75\pm0.13~, \\
  \dfrac{\upvarepsilon_{\psitwos\Kp}}{\upvarepsilon_{\jpsi 3\pip2\pim, \mathrm{NR}}} &= & 4.18\pm0.05~. 
\end{eqnarray*}
The ratio of efficiencies for the normalisation channel to that of the \btofivepi~mode is given by
\begin{equation*}
  \dfrac{\upvarepsilon_{\psitwos\Kp}}{\upvarepsilon_{\jpsi 3\pip2\pim}} = 
  k\times\dfrac{\upvarepsilon_{\psitwos\Kp}}{\upvarepsilon_{\psitwos \pip\pip\pim}} + 
  \left(1-k\right)\times\dfrac{\upvarepsilon_{\psitwos\Kp}}{\upvarepsilon_{\jpsi 3\pip2\pim,{\mathrm{NR}}}} = 5.31\pm0.06~.
\end{equation*}
The~statistical uncertainty in the ratio $k$~is accounted for in the~calculation of the~statistical uncertainty for the~ratio $R_{5\pi}$.

Since the decay products in the channels under study have similar kinematics, many 
systematic uncertainties cancel in the~ratio\,(for instance those related to muon identification). 
The~different contributions to the~systematic uncertainties affecting this analysis are described below. 
The~resulting individual uncertainties are presented in Table~\ref{table:system}. 

\begin{table}[t]
\caption{\small Relative systematic uncertainties~(in \%) for the ratios of branching fractions. The total uncertainty is the quadratic sum of the individual components.}
\begin{center}
  \begin{tabular}{p{8cm}cc}
    Source                          & $R_{\psitwos}$ & $R_{5\pi}$                  \\
    \hline
    Fit model                       & $4.6$          & $2.4$                       \\
    Decay model                     & $5.9$          & $1.1$                       \\
    Hadron interactions             & $2\times1.4$   & $2\times1.4$                \\ 
    Track reconstruction            & $1.9$          & $1.8$                       \\ 
    Hadron identification           & $0.3$          & $0.3$                       \\
    Size of the simulation sample   & $1.9$          & $1.2$                       \\
    Trigger                         & $1.1$          & $1.1$                       \\
    $\BR(\psitwos\to\jpsi\pip\pim)$ & $0.9$          &   ---                       \\ 
    \hline
    Total                           & $8.5$          & $4.7$                 
  \end{tabular}
\end{center}
\label{table:system}
\end{table}

The dominant uncertainty arises from the~imperfect knowledge of the~shape
of the~signal and the background 
in the~\Bp and \psitwos~mass distributions. 
The~dependence of the~signal yields on the~fit model is studied by varying the~signal 
and background parametrisations. 
The~systematic uncertainties are determined for the~ratios of event yields in 
different channels by taking the~maximum deviation of the~ratio 
obtained with the~alternative model 
with respect to the~baseline fit model. 
The~uncertainty determined for 
$R_{\psitwos}$ and $R_{5\pi}$ is $4.6\%$ and $2.4\%$, respectively.

To assess the systematic uncertainty related to the~\btofivepi\,(\btopsipis)~decay model 
used in the~simulation, the~reconstructed mass distribution of the~three\nobreakdash-pion\,(five\nobreakdash-pion) system 
in simulated events is reweighted to reproduce the~distribution observed in data. 
There is a~maximum change in efficiency of~$5.9\%$ for the~resonant mode 
and~$4.7\%$ for the~non\nobreakdash-resonant mode leading to a~$1.1\%$ change 
in the~total efficiency, which is taken as the~systematic uncertainty for the~decay model.

Further uncertainties arise from the differences between data and simulation, 
in particular those affecting the efficiency for the reconstruction of charged-particle tracks.
The~first uncertainty arises from the simulation of hadronic interactions in the detector,
which has an~uncertainty of~$1.4\%$ per track~\cite{LHCb-DP-2013-002}. 
Since the signal and normalisation channels differ by two tracks in the~final state, 
the~corresponding uncertainty is assigned to be $2.8\%$. 
The~small difference in the track\nobreakdash-finding efficiency between data and simulation 
is corrected using a~data\nobreakdash-driven technique~\cite{LHCb-DP-2013-002}. 
The~uncertainties in the~correction factors are propagated to 
the~efficiency ratios by means of pseudoexperiments. 
This~results in a~systematic uncertainty 
of $1.9\%$ and $1.8\%$ for the ratios of $R_{\psitwos}$ and $R_{5\pi}$, respectively. 

The uncertainties on the efficiency of hadron identification due to the~limited size of the calibration sample 
are also propagated to the efficiency ratios by means of pseudoexperiments.
The~resulting uncertainties are equal to $0.3\%$ for both branching fraction ratios.
Additional uncertainties related to the~limited size of the simulation sample are $1.9\%$ and $1.2\%$ for $R_{\psitwos}$ and $R_{5\pi}$, respectively.

The trigger is highly efficient in selecting decays with two muons in the~final state. 
The~trigger efficiency for events with a~\mbox{$\jpsi\to\mumu$} produced in beauty hadron decays 
is studied using data in high\nobreakdash-yield modes and 
a~systematic uncertainty of $1.1\%$ is \mbox{assigned} based on 
the~comparison of the~ratio of trigger efficiencies for 
high\nobreakdash-yield samples of~\mbox{$\Bu\to\jpsi\Kp$} and \mbox{$\Bu\to\psitwos\Kp$}~decays 
in data and simulation~\cite{LHCb-PAPER-2012-010}.

%% file: results.tex
\section{Results and summary}
\label{sec:results}

A~search for the~decay \btofivepi is performed using a~data sample 
corresponding to an~integrated luminosity of~$3.0\invfb$, 
collected by the~\lhcb experiment. 
A~total of $139\pm18$~signal events are observed, 
representing the~first observation of this decay channel. 
Around~half of the \Bp candidates are found to decay through 
the~\psitwos~resonance. 
The~observed yield of 
~\btopsipis~decays is $61\pm10$ events, 
which is the~first observation of this decay channel.

Using the $\Bu\to\psitwos\Kp$~decay as a normalisation channel, 
the~ratios of the~branching fractions are measured to be
\begin{align*}
  \label{eq:results}
  R_{5\pi}    &= \dfrac{\BR(\btofivepi)}{\BR(\btopsiks)} = (1.88\pm0.17\pm0.09)\times10^{-2}~, \\
  R_{\psitwos} &= \dfrac{\BR(\btopsipis)}{\BR(\btopsiks)} = (3.04\pm0.50\pm0.26)\times10^{-2}~,
\end{align*}
where the~first uncertainties are statistical and 
the~second are systematic. 
The~ratio $R_{5\pi}$ contains 
also the~contribution from $\btopsipi$~decays.

The~multipion distributions in the~\fivepi final state\,(vetoing the~\psitwos meson contribution) 
and in the~$\psitwos\pip\pip\pim$ final state are studied. 
A~structure which can be associated to the~$\Prho^{0}$~meson is seen in 
the~$\pip\pim$ combinations of the~$\fivepi$ final state. 
The~multipion distributions are compared with the~theoretical predictions from 
the~factorisation approach and a~phase\nobreakdash-space model. 
The~prediction from the~factorisation approach is found to be in somewhat better agreement 
with the~data than the prediction from the~phase\nobreakdash-space model.

%% file: acknowledgements.tex
\section*{Acknowledgements}

\noindent 
We thank A.~V.~Luchinsky for interesting discussions 
and providing the~models based on QCD factorisation for 
\mbox{$\Bu\to\jpsi3\pip2\pim$} and 
\mbox{$\Bu\to\psitwos\pip\pip\pim$}~decays.
We~express our gratitude to our colleagues in the~CERN
accelerator departments for the~excellent performance of the~LHC.
We~thank the~technical and administrative staff at the~LHCb
institutes.
We~acknowledge support from CERN and from the~national
agencies: CAPES, CNPq, FAPERJ and FINEP\,(Brazil);
NSFC\,(China);
CNRS/IN2P3\,(France);
BMBF, DFG and MPG\,(Germany);
INFN\,(Italy); 
FOM and NWO\,(The~Netherlands);
MNiSW and NCN\,(Poland);
MEN/IFA\,(Romania); 
MinES and FASO\,(Russia);
MinECo\,(Spain);
SNSF and SER\,(Switzerland); 
NASU\,(Ukraine);
STFC\,(United Kingdom);
NSF\,(USA).
We~acknowledge the~computing resources that are provided by CERN,
IN2P3\,(France),
KIT and DESY\,(Germany),
INFN\,(Italy),
SURF\,(The~Netherlands),
PIC\,(Spain),
GridPP\,(United Kingdom),
RRCKI and Yandex LLC\,(Russia),
CSCS\,(Switzerland),
IFIN\nobreakdash-HH\,(Romania),
CBPF\,(Brazil),
PL\nobreakdash-GRID\,(Poland) and OSC\,(USA).
We~are indebted to the~communities behind the~multiple open 
source software packages on which we depend.
Individual groups or members have received support from
AvH Foundation\,(Germany),
EPLANET, Marie Sk\l{}odowska\nobreakdash-Curie Actions and ERC\,(European Union), 
Conseil G\'{e}n\'{e}ral de Haute\nobreakdash-Savoie,
Labex ENIGMASS and OCEVU, 
R\'{e}gion Auvergne\,(France),
RFBR and Yandex LLC\,(Russia),
GVA, XuntaGal and GENCAT\,(Spain),
Herchel Smith Fund, The~Royal Society, Royal Commission for the~Exhibition
of 1851 and the~Leverhulme Trust\,(United Kingdom).

%% file: LHCb_Authorship_flat_02-Aug-2016.tex
\centerline{\large\bf LHCb collaboration}
\begin{flushleft}
\small
R.~Aaij$^{40}$,
B.~Adeva$^{39}$,
M.~Adinolfi$^{48}$,
Z.~Ajaltouni$^{5}$,
S.~Akar$^{6}$,
J.~Albrecht$^{10}$,
F.~Alessio$^{40}$,
M.~Alexander$^{53}$,
S.~Ali$^{43}$,
G.~Alkhazov$^{31}$
% ...

{\footnotesize \it
% ...
$ ^{5}$Clermont Universit{\'e}, Universit{\'e} Blaise Pascal, CNRS/IN2P3, LPC, Clermont-Ferrand, France\\
$ ^{6}$CPPM, Aix-Marseille Universit{\'e}, CNRS/IN2P3, Marseille, France\\
% ...
$ ^{10}$Fakult{\"a}t Physik, Technische Universit{\"a}t Dortmund, Dortmund, Germany\\
% ...
$ ^{31}$Petersburg Nuclear Physics Institute (PNPI), Gatchina, Russia\\
% ...
$ ^{39}$Universidad de Santiago de Compostela, Santiago de Compostela, Spain\\
$ ^{40}$European Organization for Nuclear Research (CERN), Geneva, Switzerland\\
% ...
$ ^{43}$Nikhef National Institute for Subatomic Physics, Amsterdam, The Netherlands\\
% ...
$ ^{48}$H.H. Wills Physics Laboratory, University of Bristol, Bristol, United Kingdom\\
% ...
$ ^{53}$School of Physics and Astronomy, University of Glasgow, Glasgow, United Kingdom\\
% ...
}
\end{flushleft}